\documentclass[a4paper,fleqn,usenatbib]{mnras}

\usepackage{newtxtext}
\usepackage[shortlabels]{enumitem}

\usepackage[T1]{fontenc}
\usepackage{ae,aecompl}

\usepackage{natbib,twoopt}
\usepackage{siunitx}
\usepackage[fleqn]{amsmath}
\usepackage{graphicx}
\usepackage{placeins}
\usepackage{xcolor}
\usepackage{multirow}
\usepackage{scalefnt}

\renewcommand{\vec}[1]{\textbf{#1}}

\title[Formation scenarios for Kepler 1625b I]{Exploring formation scenarios for the exomoon candidate Kepler 1625b I}

\author[R. A. Moraes and E. Vieira Neto]{%
  R. A. Moraes$^{1,2}$\thanks{E-mail: ricardo.moraes07@gmail.com (RAM)} and
  E. Vieira Neto$^{1}$\thanks{E-mail: ernesto@feg.unesp.br (EVN)}  \\
  $^{1}$ UNESP, Univ. Estadual Paulista - Grupo de Din\^{a}mica Orbital \& Planetologia, Guaratinguet\'{a}, CEP 12.516-410, S\~{a}o Paulo, Brazil\\
  $^{2}$ Instituto Federal de Educa\c{c}\~{a}o, Ci\^{e}ncia e Tecnologia de S\~{a}o Paulo, S\~{a}o Jos\'{e} dos Campos, CEP 12.223-201, S\~{a}o Paulo, Brazil }
\date{Accepted XXX. Received YYY; in original form ZZZ}

\pubyear{2020}

\begin{document}
\label{firstpage}
\pagerange{\pageref{firstpage}--\pageref{lastpage}}
\maketitle


\begin{abstract}
If confirmed, the Neptune-size exomoon candidate in the Kepler 1625 system will be the first natural satellite outside our Solar System. Its characteristics are nothing alike we know for a satellite. Kepler 1625b I is expected to be as massive as Neptune and to orbit at 40 planetary radii around a ten Jupiter mass planet. Because of its mass and wide orbit, this satellite was firstly thought to be captured instead of formed in-situ. In this work, we investigated the possibility of an in-situ formation of this exomoon candidate. To do so, we performed N-body simulations to reproduce the late phases of satellite formation and use a massive circum-planetary disc to explain the mass of this satellite. Our setups started soon after the gaseous nebula dissipation, when the satellite embryos are already formed. Also for selected exomoon systems we take into account a post-formation tidal evolution. We found that in-situ formation is viable to explain the origin of Kepler 1625b I, even when different values for the star-planet separation are considered. We show that for different star-planet separations the minimum amount of solids needed in the circum-planetary disc to form such a satellite varies, the wider is this separation more material is needed. In our simulations of satellite formation many satellites were formed close to the planet, this scenario changed after the tidal evolution of the systems. We concluded that if the Kepler1625 b satellite system was formed in-situ, tidal evolution was an important mechanism to sculpt its final architecture.
\end{abstract}
\begin{keywords}
  planets and satellites: formation -- 
  planets and satellites: individual (Kepler1625 b I)
 
\end{keywords}


\section{Introduction}
\label{sone}

With the increasing population of detected exoplanets, questions about satellites around these bodies, the so-called exomoons, started to be addressed \citep{Barnes-Obrien-2002, Domingos-etal-2006, Cassidy-etal-2009, Namouni-2010, Heller-etal-2014, Zollinger-etal-2017, Heller-2018, Haqq-Misra-Heller-2018}. The presence of satellite families around planets in our Solar System can provide useful information about the formation history of a planet. Also, in our Solar System for instance, the Earth's spin state is believed to be a result of the giant impact of a Mars-size object and a proto-Earth, which might have given birth to the Moon \citep{Cameron-Ward-1976}. Most commonly, the satellite's orbit characteristics is used to reconstruct the circum-planetary disc around the planet, for instance, the architecture and composition of the Galilean satellites around Jupiter indicate that the disc in which the satellites were formed had a temperature profile inversely proportional to the distance to the planet and was composed with rocky and icy material \citep{Heller-etal-2015}. Outside our Solar System, exomoons might play the same role, helping to better characterize the exoplanets.

Besides providing information about the planet's formation, the exomoons are currently pointed as favourable habitats for biological life \citep{Reynolds-etal-1987, Heller-Barnes-2013}. According to \citet{Heller-Barnes-2014} and \citet{Heller-etal-2014}, the number of exomoons in stellar habitable zones might be higher than the number of rocky planets inside the same region, such as the exomoons would be the most numerous population of habitable worlds.

Because of the aforementioned reasons, the search for exomoons is a hot-topic in astronomy, and, however not yet confirmed, several satellites candidates were proposed using different techniques of detection. \citet{Bennett-etal-2014} reported signs of what would be a sub-Earth-mass satellite orbiting a gas giant planet from a microlensing event. In \citet{Ben-Jaffel-Ballester-2014}, the authors found asymmetries in the transit light curves of the exoplanets HD 189733b and WASP-12b, which could be explained through the presence of exomoons. However, in both cases the predicted satellites would be outside the Hill sphere of their respective planet. More exomoon candidates were proposed based on a single exoplanet transit using data from CoRoT \citep{Lewis-etal-2015} and from stacked light curves of the Kepler space telescope \citep{Hippke-2015}. More recently, through analysis of transit light curves from the Kepler telescope in the system Kepler 1625, \citet{Teachey-etal-2018} announced the detection of the most plausible exomoon candidate so far.

The system Kepler 1625 is formed by a star (Kepler 1625), a planet (Kepler 1625b) and a satellite candidate (Kepler 1625b I) orbiting the planet. Kepler 1625 is a $8.7{\pm1.8}$~Gyr old \citep{Teachey-Kipping-2018} G-type star with mass $M_{star} \sim 1.079$~$M_{\odot}$ and radius $R_{star} \sim 1.793$~$R_{\odot}$, almost $2\,181$~pc distant from us \citep{Mathur-etal-2017}. Kepler 1625b is orbiting the star with a semi-major axis of $a_{p} \sim 0.87$~au \citep{Morton-etal-2016}. The mass of the planet is yet to be confirmed, however photo-dynamical fits and fits of transit light curves presented in \citet{Teachey-etal-2018}, suggested a mass of $M_p \sim 10$~$M_{Jup}$ and radius of $R_p \sim 1.18$~$R_{Jup}$ (see \citet{Heller-2018} for a more complete range of masses for Kepler 1625b). The mass and radius of the exomoon candidate are also not yet determined, but according to \citet{Teachey-Kipping-2018} these values should be similar to the ones of planet Neptune. In this way, Kepler 1625b I would be the largest and more massive satellite ever detected. Currently, the orbital separation between the exomoon and exoplanet is still poorly constrained. Firstly, \citet{Teachey-etal-2018} predicted a semi-major axis of $a_s = 19.1^{+2.1}_{-1.9}$~$R_p$, which would imply a tidally evolved satellite. However, in \citet{Teachey-Kipping-2018} the authors find that the exomoon's semi-major axis could be wider, around $40$~$R_p$, inside the planet's stability region. More recently, \citet{Martin-etal-2019} calculated the satellite-planet separation to be $a_s\sim39.9^{+15.5}_{-9.1}$~$R_p$, based on the information of \citet{Teachey-Kipping-2018}. For a canonical value, we adopt $a_s\sim40$~$R_p$ as the current location of Kepler 1625b I. A summary of the system's information is given in Table \ref{tab:properties}.

\begin{table*}
\caption[Properties]{Some physical elements of the system Kepler 1625 as given by \citet{Morton-etal-2016}, \citet{Mathur-etal-2017}, \citet{Teachey-etal-2018} and \citet{Teachey-Kipping-2018}, which have been used in the simulations.}
\begin{tabular}{cccccccc}\hline 
 $M_{star}$	    & $R_{star}$ 	& $M_p$       & $R_p$     & $a_p$    & $M_s$     & $R_s$     & $a_s$			\\
 $M_{\odot}$    & $R_{\odot}$   & $M_{Jup}$   & $R_{Jup}$ & $ua$     & $M_{Nep}$ & $R_{Nep}$ & $ R_p$ $(R_{Jup}) $        	\\ \hline	
 $1.079$        & $1.793$       & $10.0$      & $1.18$    & $0.87$   & $1.0$     & $1.0$     & $ 40.0$ $(47.2)$          \\ \hline
\end{tabular}
\label{tab:properties}
\end{table*}

Because of its peculiar characteristics, the origin of Kepler 1625b I is a challenge for theorists \citep{Teachey-Kipping-2018}. In \citet{Heller-2018}, it is discussed three possible origin scenarios for Kepler 1625b I:
\begin{itemize}
    \item impact of another body into Kepler 1625b;
    \item in-situ accretion;
    \item capture.
\end{itemize}
The author argued that the satellite-to-planet mass ratio is more than one order of magnitude larger than the scaling law factor found for our Solar System, $10^{-4}$ \citep{Canup-Ward-2006}, in this way, the mass of the satellite would not be compatible to a formation in a circum-planetary disc. Thus, the author favours a formation by capture, where the satellite is gravitationally captured by the planet and, due to tidal interaction, would migrate to its current orbital position. More information, such as the orbital direction of motion of the satellite, could be determinant for his conclusion. Following \citet{Heller-2018}, \citet{Hamers-Zwart-2018} proposed a tidal capture model for Kepler 1625b I, arguing that the satellite candidate was in fact a planet before its capture. In order for the tidal evolution to be effective, the capture must have occurred early in the evolution of the planetary system, indicating that the exomoon has been orbiting the planet for more than a Gyr. Also, the authors postulated that this kind of capture is not uncommon and should be a trend in extrasolar systems.

Here, we argued that, just like for exoplanets, the patterns and scale laws found for our Solar System might not apply to exomoons, since the environments where these bodies could be found are extremely different from what we have in our system. In particular, the argument about the satellite-to-planet mass ratio of $10^{-4}$ might not be applied at all outside our Solar System since this value is highly dependent in many specific parameters as we can see in equation 2 of \citet{Canup-Ward-2006}. In this way, we do not see how such an argument could be used to discard the in-situ formation scenario. Thus, we revisited models of formation in a circum-planetary disc, testing the plausibility of formation of exomoons in very massive solid-rich discs, which would be the case for Kepler 1625b I. 

To validate the assumption of in-situ formation, one might have to argue about how the amount of mass needed for satellite formation was delivered to the circum-planetary disc. There is a common sense that during the formation of giant planets, these bodies became so massive that a gap was carved into the proto-planetary disc, separating the proto-planet from the rest of the disc. However, as shown by \citet{Kley-1999} and \citet{lubow}, even with a gap the circum-planetary disc and the proto-planetary disc were still connected by ``spiral arms'', in such a way that material could flow through these ``arms'' towards the planet in form of gas and fine dust. Then the question becomes, would this fine dust be enough to produce a massive circum-planetary disc? According to \citet{Sasaki-etal-2010} and \citet{Ida-Lin-2004} this hypothesis is very unlikely, because once the gap is opened the influx of solid material from the proto-planetary disc dramatically  decreases, and also at the moment of the gap opening the proto-planetary disc would be already poor in solids. We argue that most of the solids presented in the circum-planetary disc are formed by leftover bodies that were not accreted by the planet, dragged into the planet's orbit before the formation of the gap. As pointed out by \citet{Barr-2016} during its formation a more massive planet, such as Kepler 1625b, will liberate more gravitational energy during its contraction phase, creating a hotter circum-planetary disc, in this way more solid material would be attracted into planetary orbit and massive satellites could potentially be formed. Also, according to  \citet{Szulagyi-2017} the mass in a circum-planetary disc scales not only with the mass of the planet, but also with the mass of the proto-planetary disc. Thus, by assuming the circum-planetary disc around Kepler 1625b to be very massive, we are assuming that the whole proto-planetary disc is massive and heavily composed by solids.

Under the above assumption, we explore the in-situ accretion models focusing on the amount of solid material needed to a Neptune-like body to form around a $10$ $M_{Jup}$ planet. In our models, we considered solid-enhanced massive discs \citep{Mosqueira-Estrada-2003a, Mosqueira-Estrada-2003b} after the dissipation of gaseous portion of the circum-planetary disc, which means that the embryos core was already formed and migrated inwardly into the stability zone of the planet. We performed several N-body simulations using the package MERCURY \citep{Chambers-1999} to analyse regions where particles are stable and to follow the formation of satellite systems around Kepler 1625b and their tidal evolution.

Besides the scenarios presented in \citet{Heller-2018} and the one presented here, It is also possible the scenario of accretion from rings to form this satellite \citep{Charnoz-etal-2011, Crida-Charnoz-2012}. This scenario is based in a massive ring of particles that spreads out of the Roche limit of the planet while the satellites are formed following an hierarchical pattern for their mass distribution, with the farthest satellite being the more massive. This model was successfully applied to explain the formation of the mid-sized satellites of Saturn and Uranus orbiting close to their host planet. If the exomoon candidate is confirmed at the location firstly predicted in \citet{Teachey-etal-2018} this model must be considered and, in this case, the mass initially in the ring surrounding Kepler 1625b would be an object of study.

This paper is organized as follows. In Section \ref{stwo}, we detail our models, explain the parameters used and study the stability of the system. The results of the simulations
are presented in Section \ref{sthree}, and in Section \ref{conc}, we draw our conclusions.

\section{Models}\label{stwo}

It is generally believed that gaseous giant planets were formed farther away from the star, in regions where gas and solids were more abundant, and then they experienced a migration inwardly through type I and type II migration regimes \citep{Nelson-etal-2000,Hamers-etal-2016}. While planets migrate inwardly their Hill sphere shrinks, which could lead to the loss of satellites by ejection or collision with the host planet \citep{Namouni-2010}. In this work, in order to cover the formation of satellites in different stages of planetary evolution, we investigated the formation of satellites in the system Kepler 1625 considering four different star-planet separations, $0.87$~au (current separation), $1$~au, $1.75$~au, and $2$~au (after this distance the results became very similar).

This work is divided in three dependable phases: 
    \begin{enumerate}[a)] 
        \item The study of stability around Kepler 1625b; 
        \item The evolution of satellite embryos around the planet; 
        \item The tidal evolution of the surviving satellites.
    \end{enumerate}
In this section we will present the analysis regarding the stability of the system and the numerical models for phases b) and c).

\subsection{Stability}

In order to search for regions of stability around Kepler 1625b, we distributed 10\,000 massless particles inside the Hill radius of the planet. We used a planetocentric coordinates system with the star performing a circular motion around the planet with four different star-planet separations. For our purposes, the particles were exposed only to gravitational forces from the planet and the star, and they do not interact with each other.

In our disc, particles had their semi-major axes randomly distributed from the arbitrary small distance of $5.0$~$R_p$ to $1.0$~$R_H$ from the planet. Other orbital elements were initially set to zero, except the mean longitude which was randomly taken between $0^{\circ}$ and $360^{\circ}$. Damping effects such as gas drag were not considered and the simulations were performed for 10\,000 years.

It is important to say that the Hill radius depends on the separation star-planet, as we can see in equation
\begin{align}\label{eq:hill}
    R_H = a_p\left(\dfrac{M_{p}}{3M_{star}}\right)^{1/3},
\end{align}
where $a_p$ is the semi-major axes for each separation. Thus, for each separation the outer distance had a different measure in $R_p$.

\begin{figure*}
    \begin{center}
        \includegraphics[scale=.3]{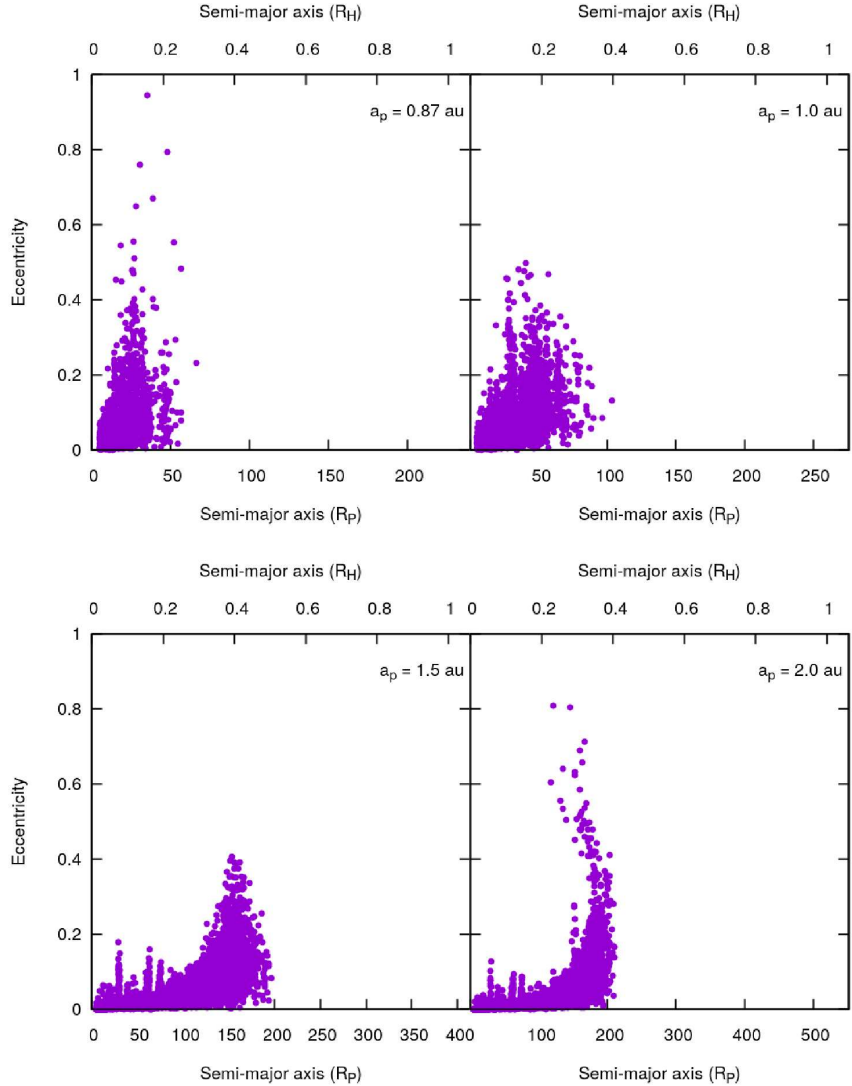}
        \caption[Density.]{These diagrams of semi-major axis versus eccentricity show the results of a dynamical evolution of a disc composed with massless particles around Kepler 1625b with star-planet separations of $a_p=0.87$~au, $a_p=1$~au, $a_p=1.5$~au, $a_p=2$~au. The upper axis shows the distance in $R_H$ and the lower axis in $R_p$.}
        \label{fig:stability} 
    \end{center} 
\end{figure*}

In Fig. \ref{fig:stability} we show the results of the dynamical evolution of the particles for $10\,000$ years with separations of $a_p=0.87$~au, $a_p=1$~au, $a_p=1.5$~au, $a_p=2$~au, respectively. The upper axis shows the distance in $R_H$ and the lower axis in $R_p$.

As expected, the star plays a major role in the stability of the system. In all figures, observing the upper axis with the measure in Hill's radius, one can see that the gravitational forces arising from the star are strong enough to clear more than $50$ $\%$ of the region where the disc was distributed. Particles in these regions were destabilized by a resonance known as evection resonance \citep{Yokoyama.etal-2008, Vieira-Neto-etal-2006}. The evection resonance is responsible for an excitation on the particles' eccentricity leading to scattering after some time \citep{Hamilton-Krivov-1997}. In satellite capture studies, usually it is considered as gravitationally stable, regions inside half of the Hill radius of the planet, $a_{cap}< 0.5$~$R_H$, \citep{Vieira-Neto-etal-2006, Hamers-Zwart-2018}. Our simulations show that this outer limit may look to be overestimated for Kepler 1625b, however it is due to our planar and circular initial conditions.

For setup with $a_p = 0.87$~au (upper left panel of Fig. \ref{fig:stability}), we see that the particles are stable only within around $0.22~R_H$ ($\sim 50$ $R_p$), well inside the maximum semi-major axis for prograde satellites proposed by \citet{Domingos-etal-2006}. Also, we found that they present eccentric configurations, since particles are confined in a tight region. In this way, we can expect to find satellites with eccentric orbits when they are locally formed in such small discs.

Increasing the separation between star and planet we see the gravitational influence of the star being less disruptive and showing bodies with lower eccentricity in regions above 0.2~au. We can see from setups with $a_p=1$~au to $a_p=1.5$~au that the outer boundary of stability is pushed farther in the disc, and the particles become less eccentric (except for the ones located close to the outer boundary). In these cases the evection resonance appears to be less effective in the inner region, but still analysing panels with separation $a_p = 1$~au and $a_p=1.5$~au of Fig. \ref{fig:stability} we can see signs of particles' scattering near the outer boundary. Also, the outer boundary limit follows the results of \citet{Vieira-Neto-etal-2006} and \citet{Hamers-Zwart-2018} being located at $\sim 0.38458$~$R_H$ and $\sim 0.51278$~$R_H$ for setups with $a_p=1$~au and $a_p=1.5$~au, respectively.

Set up with $a_p=2$~au (lower right panel of Fig. \ref{fig:stability}) may appear to have some odd outcomes. As the separation between planet and star was increased and, as a consequence, its Hill sphere expanded, it was expected that the influences of the star to be less significant and a bigger portion of the disc should be stable. Our results show a region similar in size with the one from setup with $a_p=1.5$~au, which could mean that, when compared to the Hill radius of the planet, the region of stability shrunk. However the eccentricity of the particles more distant from the planet had increased exponentially. It is possible to observe the same pattern in setup with $a_a=1.5$~au, but yet, in this model, the growth rate of the eccentricities is higher than what we previously saw, which is not intuitive. This happen due to the evection effects which acts on the particles near the outer boundary and the survivors have high eccentricities \citep{Yokoyama.etal-2008}.

From the results regarding stability in the disc we can estimate an inner and outer boundaries ($R_{out}$) for the disc where particles are stable. We estimate for the outer boundary $50$~$R_p$ for separation $a_p=0.87$~au, $100$~$R_p$ for separation $a_p=1$~au, and $200$~$R_p$ for separations $a_p=1.5$~au and $a_p=2$~au, as one can see on the lower axis of Fig. \ref{fig:stability}. These results will be used to describe the limits of the circum-planetary disc for satellite formation purposes.

\subsection{Formation of the Satellites}

After studying the stability in the disc region, we shall move to the description of the circum-planetary disc for the formation of the exomoons candidates.

Our coordinate system will be centred on the planet and the effects of the star will be considered as an external force, thus we will not integrate the orbital evolution of the star, but consider the body in a circular keplerian motion around the planet. The equations of motion of a embryo $k$ at distance $r_k$ from the central planet are,
\begin{align}\label{eq:motion1}
&\dfrac{d^2\vec{r}_k}{dt^2} = -GM_p\dfrac{\vec{r}_k}{|\vec{r}_k|^3} - \sum\limits_{i\neq k}GM_i\dfrac{\vec{r}_k-\vec{r}_i}{|\vec{r}_k-\vec{r}_i|^3} - \sum\limits_i GM_i\dfrac{\vec{r}_i}{|\vec{r}_i|^3}\nonumber \\
& - \sum\limits_{j}GM_j\dfrac{\vec{r}_k-\vec{r}_j}{|\vec{r}_k-\vec{r}_j|^3} - \sum\limits_j GM_j\dfrac{\vec{r}_j}{|\vec{r}_j|^3} - GM_{star}\dfrac{\vec{r}_{star}}{|\vec{r}_{star}|^3}\nonumber \\
& - GM_{star}\dfrac{\vec{r}_k-\vec{r}_{star}}{|\vec{r}_k-\vec{r}_{star}|^3}, 
\end{align}
where $k = 1, 2, \ldots, n_e$ and $i$ are the satellite embryos in the disc, $j$ are the satellitesimals in the disc, $G$ is the gravitational constant, $r_{i}$ is the distance planet-embryo, $r_{j}$ is the distance planet-satellitesimal, $r_{star}$ is the planet-star distance, $M_i, M_j, M_p, M_{star}$ are the masses of the embryo $i$, of satellitesimal $j$, of the central planet, and the star, respectively. The terms on the right side of Eq. \ref{eq:motion1} are the gravitational force from the planet on embryo $k$, the mutual gravity interaction between the embryos and its indirect terms, the gravitational interaction between the embryos and the satellitesimals and its indirect terms, the indirect terms from the star and the gravitational interaction with the star, respectively. We have a similar equation for the satellitesimals with the difference that they interact with the embryos, but they do not interact with each other.

Our work aims to establish constraints on the amount of solids on the disc in which Neptune-like satellites could form. We are considering that these solids come from ice reach particles. The interactions of solids with a possible remaining gas just after the formation of satellites were neglected. Post-formation tidal evolution will be addressed later in the manuscript.

As in the study of stability, we will consider four different star-planet separations for our simulations. And follow the formation of the satellites for 100\,000 years.

\subsection{Solid Disc}

The solid disc will be composed by massive embryos and less massive satellitesimals. Following \citet{Moraes-etal-2018}, the embryos will gravitationally interact with each other and with the satellitesimals, while the interactions between satellitesimals are neglected.

Since we are interested in finding the necessary amount of solid materials in the disc just enough for the formation of a Kepler 1625b I-like exomoon, we will simulate circum-planetary discs containing from $1$ to $6$ Neptune's masses in solids. In order to decrease the dependence on the initial distribution of the embryos, for each disc with a given mass, we simulate ten cases randomly distributing the bodies.
    
In all our setups we will consider 40 embryos and 2\,000 satellitesimals. As in the works of \citet{Kokubo-Ida-2002} and \citet{Raymond-etal-2005}, the masses of the embryos are proportional to $r^{3/4}$, in order to have the more massive bodies farther in the disc, while in the inner parts of the disc it will be populated by the smaller bodies. All satellitesimals have the same mass and are uniformly distributed throughout the disc.

As we are simulating an in-situ formation process for the satellites, we will consider the collisions between embryos and between embryos and satellitesimals to have a relative velocity in such a way that the accretion is possible. After the collision the two bodies will inelastically merge to produce a new body with conservation of mass and linear momentum, satellitesimals vanishes and embryos grows. The central body accrete the embryos and satellitesimals, while collisions between satellitesimals are neglected.

Based on our findings about stability, it is not necessary to consider a wide circum-planetarydisc, extending for the whole Hill radius of the planet. Thus, we choose the inner and outer boundaries of the circum-planetary disc to follow the limits of stability found before. For each separation planet-star we have six discs from 1 to 6 Neptune masses. Each disc has its own solid distribution based on its mass quantity.

For our purposes, initially all the bodies have circular orbits and inclinations $<10^{-4}$ (to allow inclination to increase/decrease during the simulation). The mean longitudes are randomly taken between $0^{\circ}$ and $360^{\circ}$. The other two angular orbital elements, node and pericentre, were set to zero.

A summary of all setups is shown in Table \ref{tab:formation}, which display all relevant parameters. Every disc has its initial inner boundary at 5~$R_p$.

\begin{table*}
\caption[Formation]{Parameters of the simulations regarding the satellites formation. Here, $a_p$ is the semi-major axis of the planet, the separation star-planet; $M_{emb}$ is the range of mass for the embryos; $M_{satel}$ is the mass of each satellitesimal; $R_{out}$ is the initial outer boundary of the particle's disc and it was obtained from the stability studies.}
\begin{tabular}{lcccc}\hline 
 Set up	              & $a_p$  &$M_{emb}$       &$M_{satel}$&$R_{out}$		\\
                      & au     &$\times10^{-2}M_{Nep}$      & $\times10^{-4}M_{Nep}$       & $R_p$       \\ \hline	 
 \texttt{kepler-087-1}&\multirow{6}{*}{$0.87$}&$0.34 - 1.91$&$2.52$& \multirow{6}{*}{$50.0$} 	\\ 	
 \texttt{kepler-087-2}&        &$0.68 - 3,82$   & $5.05$   &    \\ 
\texttt{kepler-087-3} &        &$1.00 - 5.63$   & $7.57$   &    \\ 
\texttt{kepler-087-4} &        &$1.36 - 7.64$   & $10.10$  &    \\ 
\texttt{kepler-087-5} &        &$1.70 - 9.55$   & $12.62$  &    \\ 
\texttt{kepler-087-6} &        &$2.04 - 11.50$  & $15.15$  &    \\ \hline
\texttt{kepler-1-1}   &\multirow{6}{*}{$1.0$}&$0.26 - 2.47$&$2.52$& \multirow{6}{*}{$100.0$}   \\
\texttt{kepler-1-2}   &        &$0.52 - 4.94$   & $5.05$   &   \\
\texttt{kepler-1-3}   &        &$0.78 - 7.41$   & $7.57$   &   \\
\texttt{kepler-1-4}   &        &$1.05 - 9.88$   & $10.10$  &   \\
\texttt{kepler-1-5}   &        &$1.31 - 12.35$  & $12.62$  &   \\
\texttt{kepler-1-6}   &        &$1.57 - 14.82$  & $15.15$  &   \\ \hline
\texttt{kepler-15-1}  &\multirow{6}{*}{$1.5$}&$0.13 - 2.08$& $2.52$& \multirow{6}{*}{$200.0$}   \\
\texttt{kepler-15-2}  &        &$0.26 - 4.15$   & $5.05$   &   \\
\texttt{kepler-15-3}  &        &$0.39 - 6.23$   & $7.57$   &   \\
\texttt{kepler-15-4}  &        &$0.52 - 8.31$   & $10.10$  &   \\
\texttt{kepler-15-5}  &        &$0.65 - 10.39$  & $12.62$  &   \\
\texttt{kepler-15-6}  &        &$0.78 - 12.47$  & $15.15$  &   \\ \hline
\texttt{kepler-2-1}   &\multirow{6}{*}{$2.0$}&$0.13 - 2.08$&$2.52$& \multirow{6}{*}{$200.0$}   \\
\texttt{kepler-2-2}   &        &$0.26 - 4.15$   & $5.05$   &   \\
\texttt{kepler-2-3}   &        &$0.39 - 6.23$   & $7.57$   &   \\
\texttt{kepler-2-4}   &        &$0.52 - 8.31$   & $10.10$  &   \\
\texttt{kepler-2-5}   &        &$0.65 - 10.39$  & $12.62$  &   \\
\texttt{kepler-2-6}   &        &$0.78 - 12.47$  & $15.15$  &   \\ \hline

\end{tabular}
\label{tab:formation}
\end{table*}

The surviving embryos will be the satellites formed through the collisional process mentioned above after 100\,000 years. At this time, the solid disc was almost completely accreted and/or depleted, and there are not many bodies left to interact with the satellites that were formed in the process, thus they should not dramatically change their orbital characteristics.

\subsection{Tidal Evolution}

After the formation of the satellite systems, we will analyse each case and select the systems with at least one satellite with masses between $0.75$ and $1.25$ $M_{Nep}$ to study the tidal evolution of these bodies. In this phase we will neglect the effects of the star and focus mainly on the contribution of tides for the final architecture of the satellite systems. In addition, we will consider the contribution of a possible oblateness of Kepler 1625b $(J_{2,p})$.

We will follow \citet{Mignard-1979} and \citet{Hussmann-etal-2019} and parametrize the tidal force by a constant time lag. In a coordinate system centred on the planet, the equation of motion of a satellite $k$ with a mass $M_k$ at distance $r_k$ from the central planet is given by,
\begin{align}\label{eq:motion2}
&\dfrac{d^2\vec{r}_k}{dt^2} = -GM_p\dfrac{\vec{r}_k}{|\vec{r}_k|^3} - \sum\limits_{i\neq k}GM_i\dfrac{\vec{r}_k-\vec{r}_i}{|\vec{r}_k-\vec{r}_i|^3} - \sum\limits_i GM_i\dfrac{\vec{r}_i}{|\vec{r}_i|^3}\nonumber \\
&+\dfrac{\left(M_p+M_k\right)}{M_pM_k}\left(\vec{f}_k - \vec{f}_{p,k}+\vec{g}_k\right) + \sum\limits_{i\neq k}\dfrac{\vec{f}_i - \vec{f}_{p,i}+\vec{g}_i}{M_P},
\end{align}
where $f_i$ and $f_k$ are the tidal forces on the satellites, $f_{p,(k,i)}$ is the tidal force on the planet due to the satellites, and $g_{(k,i)}$ is the force due to the oblateness of the planet on the satellites.
        
The expressions for the tidal forces $f_{(i,k)}$ are given in \citet{Mignard-1979} and \citet{Hussmann-etal-2019} as:
\begin{align}\label{eq:fi}
 \vec{f}_i = -3\kappa_{2,i}\Delta t_i\dfrac{GM_p^2R_i^2}{r_i^{10}}\left[2\vec{r}_i\left(\vec{r}_i\cdot \vec{v}_i\right)+r_i^2\left(\vec{r}_i \times \mathbf{\Omega}_i+\vec{v}_i\right)\right],
\end{align}
where $\Omega_i$ is the angular velocity of the satellites, $\kappa_{2,i}$ is the second order Love number and $\Delta t_i$ is the time lag. For the product of the second order Love number by the time lag we follow \citet{Bolmont-etal-2015} and assume $\kappa_{2,i}\Delta t_i = 213$~s, the same value of Earth \citep{Neron-Laskar-1997}. The radius of the satellites, $R_i$, were calculate using the numerical fit presented in \citet{Fortney-etal-2007}
\begin{align}\label{eq:fortney}
 &R_i = \left(0.0912 f_{ice}+0.1603\right)\left(\log M_i\right)^2\\ \nonumber
 &+ \left(0.3330 f_{ice}+0.7387\right)\left(\log M_i\right)\\ \nonumber
 &+\left(0.4639 f_{ice}+1.1193\right)
\end{align}
with $f_{ice}$ being the ice mass fraction related with the composition of the body ($1.0$ for pure ice and $0.0$ for pure rock). Here we choose $f_{ice} = 0.5$.

The tidal forces on the planet are given by the sum over the individual forces raised by each satellite, $\vec{f}_p = \sum \limits_i \vec{f}_{p,i}$. The tidal force induced by the $i$-th satellite is 
\begin{align}\label{eq:fpi}
 \vec{f}_{p,i} = 3\kappa_{2,p}\Delta t_p\dfrac{GM_i^2R_p^2}{r_i^{10}}\left[2\vec{r}_i\left(\vec{r}_i\cdot \vec{v}_i\right)+r_i^2\left(\vec{r}_i \times \mathbf{\Omega}_p+\vec{v}_i\right)\right],
\end{align}
where $\mathbf{\Omega}_p$ is is the angular velocity of the planet and $\kappa_{2,p}\Delta t_p$ is the product of the second order Love number by the time lag for the planet, given by \citep{Bolmont-etal-2015}
\begin{align}\label{eq:love-planet}
 \kappa_{2,p}\Delta t_p = \dfrac{3R_p^5\sigma_p}{2G},
\end{align}
with the dissipation factor $\sigma_p=2.006\times10^{-60}$~ g$^{-1}$cm$^{-2}$s$^{-1}$.

The forces arising from the oblateness of the planet are given by \citep{Beutler-2005, Hussmann-etal-2019},
\begin{align}\label{eq:gi}
 \vec{g}_i = -\dfrac{3G\left(M_p+M_i\right)R_p^2}{2r_i^5}J_{2,p}\vec{r}_i^5,
\end{align}
where the rotation deformation of the planet is described by the parameter $J_{2,p}$, with $\kappa_{2,p}=0.379$ \citep{Bolmont-etal-2015},
\begin{align}\label{eq:J2}
 J_{2,p} = \kappa_{2,p}\dfrac{\Omega_p^2R_p^3}{3GM_p}.
\end{align}
The description of the systems simulated under the assumption mentioned above will be given in the section \ref{sthree}.

\section{Results}\label{sthree}

In this section we present results obtained for the satellites formation and their tidal evolution using models described in the last section. Ours simulations treat discs with 1 to 6 Neptune masses in solids, with embryos and satellitesimals radially distributed in discs with different sizes, according to the separation between the star and planet. In an attempt to minimize the effects of the initial distribution over our final results regarding satellite formation, we simulate $10$ cases with different random distribution of solids for each of our sets, because of this we will generally discuss average results for the semi-major axis, mass, eccentricity and inclination of the formed satellites, individual cases will be spotted when necessary. For the tidal evolution simulations, it was selected systems with at least one satellite with mass between $0.75$ and $1.25$ $M_{Nep}$. The description of the selected systems and the results will be presented in sub-section \ref{subsec:tides}. 

\subsection{Satellite Formation}

In Table \ref{tab:results-formation} we present the simulation results for satellite formation. As mentioned before, because for each setup it was performed ten simulations, we present in this table average results of each setup showing the number of satellites formed and their final orbital elements. We highlighted in Table \ref{tab:results-formation}, in column Favourable, the number of satellite systems in each setup selected to be included in the tide simulations. 

\begin{table*}
\caption[Results Formation]{Results of the simulations regarding satellites formation. Here, $\overline{n}_{sat}$, is the average number of satellites formed in each model; $\overline{a}_{sat}$, $\overline{M}_{sat}$, $\overline{e}_{sat}$ and $\overline{I}_{sat}$ are average semi-major axis, mass, eccentricity and inclination of the formed satellites in each model; Column Favourable refers to the number of systems with at least one satellites with mass between $0.75$ and $1.25$ $M_{Nep}$ .}
\begin{tabular}{lcccccc} \hline
 Model	                 & $\overline{n}_{sat}$ & $\overline{a}_{sat}$ & $\overline{M}_{sat}$ & $\overline{e}_{sat}$ & $\overline{I}_{sat}$ & Favourable\\	
                         &                      & $R_{p}$              & $M_{Nep}$            &                      & Degrees              &       \\  \hline
 \texttt{kepler-087-1}   & $2.5$                & $21.68$              & $0.35$               & $0.15$               & $1,47$ 	            & 1     \\ 	
 \texttt{kepler-087-2}   & $2.4$                & $19.93$              & $0.66$               & $0.19$               & $2,23$               & 7     \\ 
\texttt{kepler-087-3}    & $2.0$                & $23.85$              & $1.17$               & $0.17$               & $0,96$               & 5     \\ 
\texttt{kepler-087-4}    & $1.9$                & $20.26$              & $1.67$               & $0.18$               & $0,97$               & 1     \\ 
\texttt{kepler-087-5}    & $1.7$                & $21.98$              & $2.15$               & $0.18$               & $0,67$               & 2     \\  
\texttt{kepler-087-6}    & $1.33$               & $23.56$              & $2.06$               & $0.27$               & $0,72$               & 2     \\  \hline
\texttt{kepler-1-1}      & $2.8$                & $30.91$              & $0.25$               & $0.13$               & $2.51$ 	            & 0     \\  	
\texttt{kepler-1-2}      & $2.0$                & $27.12$              & $0.61$               & $0.20$               & $3.17$               & 4     \\ 
\texttt{kepler-1-3}      & $1.8$                & $21.15$              & $0.90$               & $0.22$               & $5.02$               & 4     \\ 
\texttt{kepler-1-4}      & $1.9$                & $25.42$              & $1.13$               & $0.19$               & $4.21$               & 5     \\ 
\texttt{kepler-1-5}      & $1.8$                & $23.13$              & $1.67$               & $0.19$               & $3.20$               & 2     \\ 
\texttt{kepler-1-6}      & $1.9$                & $27.82$              & $1.81$               & $0.17$               & $1.55$               & 3     \\ \hline
\texttt{kepler-15-1}     & $3.1$                & $45.61$              & $0.13$               & $0.16$               & $4.27$               & 0     \\ 	
\texttt{kepler-15-2}     & $2.3$                & $32.83$              & $0.27$               & $0.13$               & $5.82$               & 0     \\ 
\texttt{kepler-15-3}     & $2.6$                & $41.61$              & $0.43$               & $0.17$               & $6.58$               & 5     \\ 
\texttt{kepler-15-4}     & $2.0$                & $40.11$              & $0.60$               & $0.25$               & $6.41$               & 6     \\ 
\texttt{kepler-15-5}     & $1.9$                & $26.82$              & $0.74$               & $0.17$               & $6.69$               & 5     \\ 
\texttt{kepler-15-6}     & $2.1$                & $41.41$              & $0.99$               & $0.22$               & $6.92$               & 6     \\ \hline
\texttt{kepler-2-1}      & $3.1$                & $53.92$              & $0.14$               & $0.16$               & $5.19$           	& 0     \\ 	
\texttt{kepler-2-2}      & $2.7$                & $44.35$              & $0.31$               & $0.16$               & $6.54$               & 0     \\ 
\texttt{kepler-2-3}      & $2.0$                & $41.10$              & $0.50$               & $0.25$               & $4.99$               & 2     \\ 
\texttt{kepler-2-4}      & $2.6$                & $51.95$              & $0.67$               & $0.21$               & $6.58$               & 7     \\ 
\texttt{kepler-2-5}      & $1.8$                & $40.09$              & $0.93$               & $0.22$               & $8.69$               & 5     \\ 
\texttt{kepler-2-6}      & $2.0$                & $44.30$              & $1.01$               & $0.25$               & $5.14$               & 7     \\ \hline

\end{tabular}
\label{tab:results-formation}
\end{table*}

To allow satellite inclination to evolve, we set the initial inclination of the embryos with a small non-zero value ($10^{-4}$). Because no damping effects were added to the simulations, the satellites became inclined in the formation process as we can see from Table \ref{tab:results-formation}, with setups \texttt{kepler-2} and \texttt{kepler-15} having, in average, the most inclined satellites. This result agrees with the findings of \citet{Pu-Lai-2018} which show that spread systems are more susceptible to inclination excitation by an external perturber, in our case the star.

We can see in setups \texttt{kepler-087} and \texttt{kepler-1} a slight decrease of orbit inclination with respect to the initial amount of mass of the disc. Thus it appears to be relevant only for setups with tight configurations. According to \citet{Sotiriadis-etal-2017} the higher the initial mass in the disc is the smaller should be the inclination of the bodies formed.

In the following subsections we depict the results obtained in this table for other parameters. 

\subsubsection{Number of Satellites}

The first correlation we see is the average number of satellites formed and the initial mass in the disc. For all systems, there is a pattern indicating a slight decline in the average number of satellites as the mass in the disc increases. This correlation happens due the fact that the number of bodies did not change for any simulation. Then, for more massive discs, the mass of the bodies increases leading to more extreme systems. Close encounters in these systems tend to be more energetic and disruptive forming less satellites. 

\begin{figure}
    \begin{center}
        \includegraphics[scale=0.67]{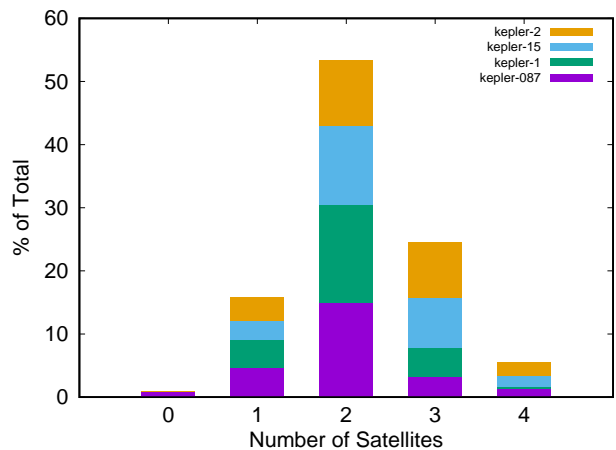}
        \caption[Number]{Percentage of cases a certain number of satellites has formed. The different colours in the bars indicate which setup satellites were formed.}
        \label{fig:number} 
    \end{center} 
\end{figure}

So far, only one satellite is predicted to orbit in Kepler 1625 system, however in the majority of our simulations more than one satellite was formed. Overall, the maximum number of satellites formed in one same system was four, while systems with no satellite were obtained only twice. In Fig. \ref{fig:number} we show the percentage of cases a certain number of satellites has formed. This figure shows that systems with two satellites are more likely and it is more common when the planet is closer to the star (setup \texttt{kepler-087}). This behaviour can be explained by the size of the circum-planetary disc. For lager separation we have larger discs (see table \ref{tab:formation}), and satellites could be formed in farther regions leaving enough distance for the formation of more populated systems. We also can see in this figure that systems with three and four satellites are more abundant in setups \texttt{kepler-15} and \texttt{kepler-2}.

It is important to notice that, besides the small disc in setup \texttt{kepler-087}, systems with four satellites were possible, in this case the surviving bodies were all very small, and usually with one satellite in the inner border and other in the outer border.

\subsubsection{Location of the Satellites}

\begin{figure*}
  \begin{center}
  \includegraphics[scale=0.67]{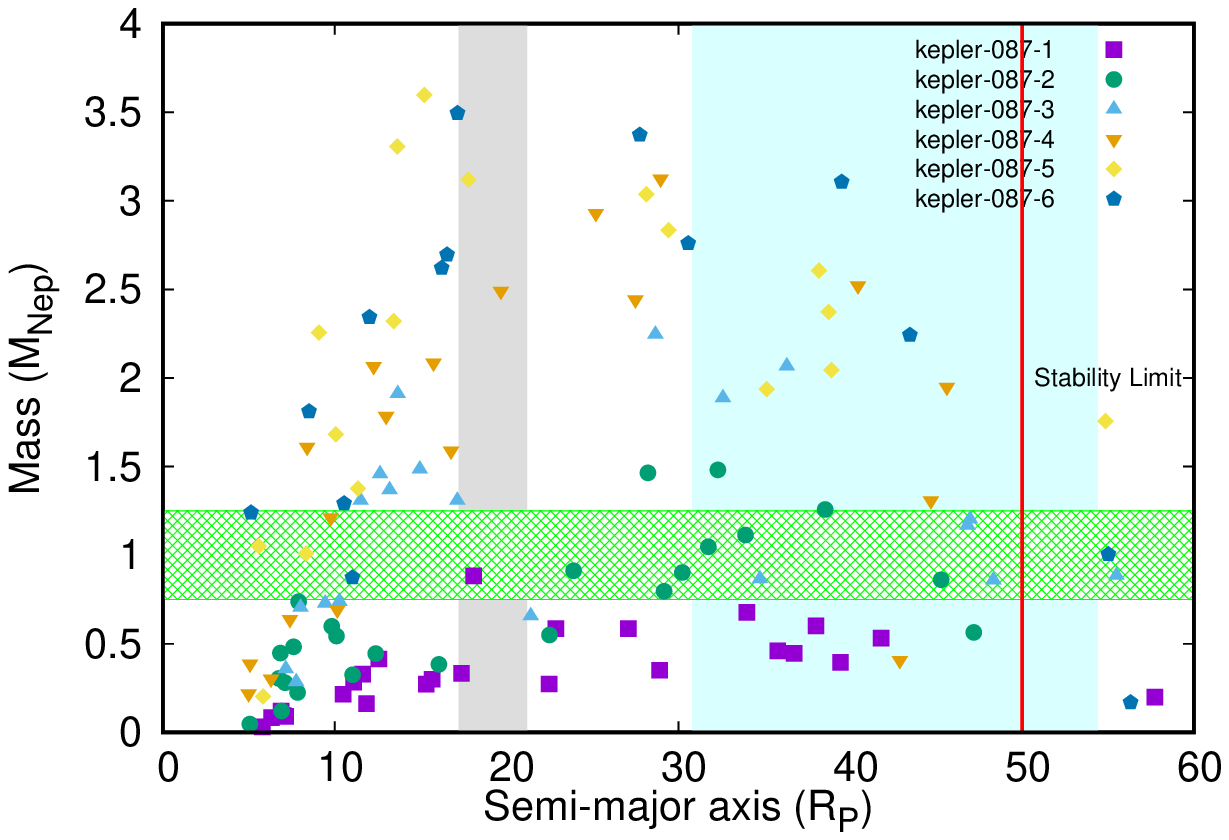}
  \includegraphics[scale=0.67]{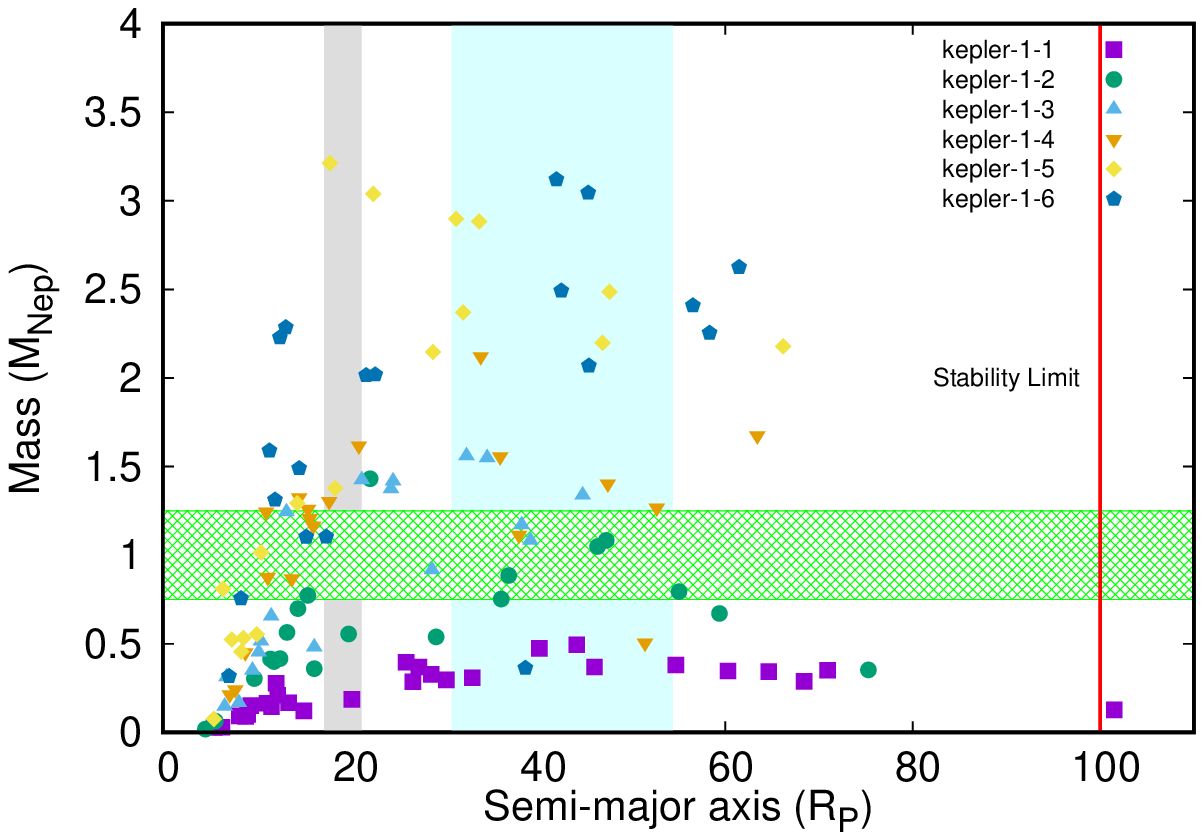}
  
  \includegraphics[scale=0.67]{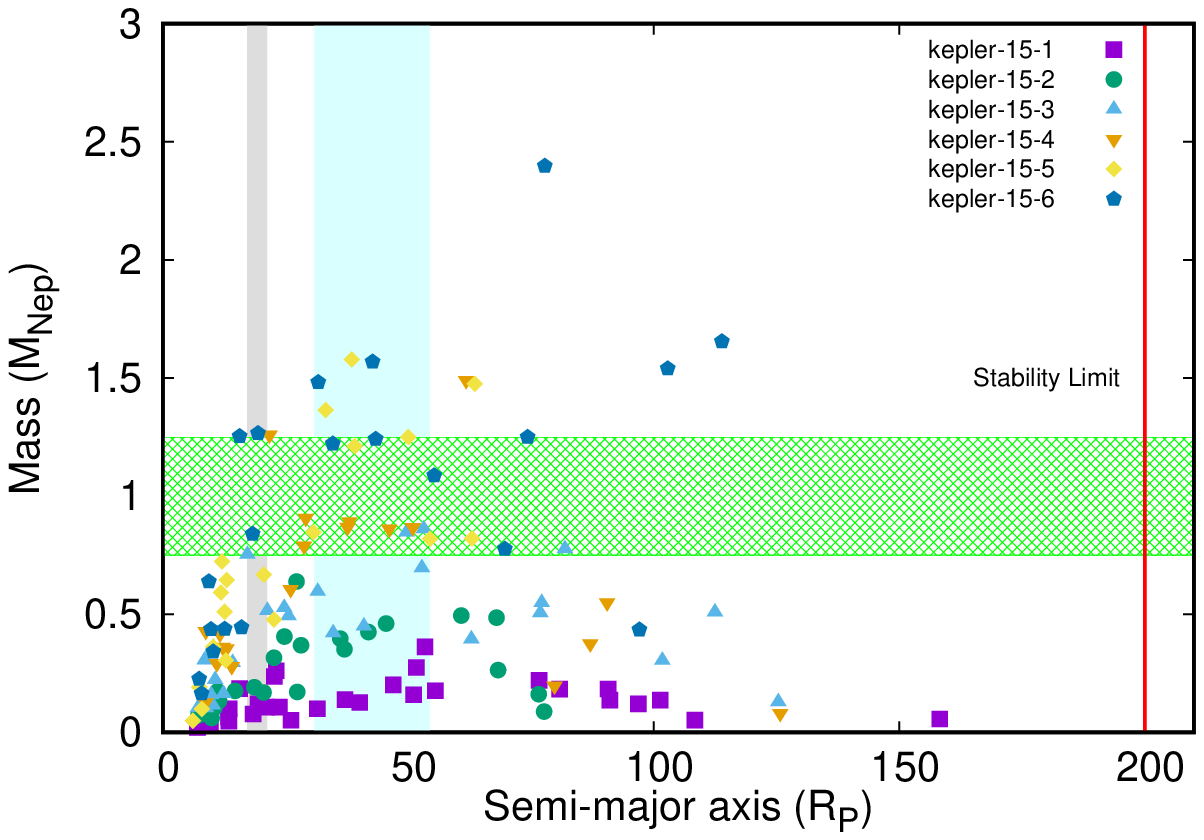}
  \includegraphics[scale=0.67]{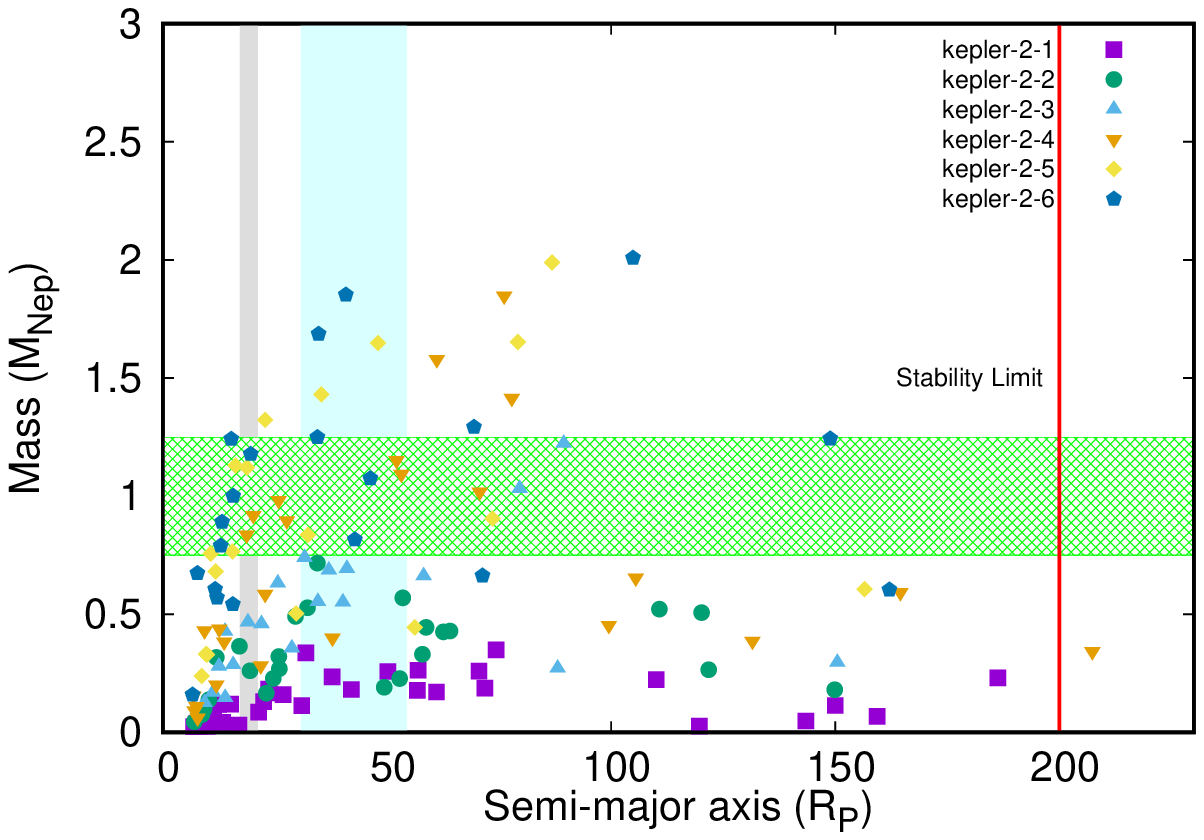}
  \caption[Kepler1]{Panels showing mass of satellites formed vs its final semi-major axis for each setup. The region in grey is the location of the exomoon according to \citet{Teachey-etal-2018} while the region in cyan is the range of semi-major axes proposed by \citet{Martin-etal-2019}. The region in green is the range of mass from $0.75$ to $1.25$~$M_{Nep}$. The red line marks the limit of stability obtained in section \ref{stwo}. The formed satellites are separated by the initial mass of the disc: 1~$M_{Nep}$ in purple squares; 2~$M_{Nep}$ in green circles; 3~$M_{Nep}$ in light-blue upward triangles; 4~$M_{Nep}$ in orange downward triangles; 5~$M_{Nep}$ in yellow diamonds; 6~$M_{Nep}$ in dark-blue pentagons.}
  \label{fig:kepler-a-m} 
  \end{center} 
\end{figure*}

In Fig. \ref{fig:kepler-a-m} we show panels for each setup star-planet separation with the mass of all formed satellites according to their distance from the planet. Region in grey is the location of Kepler 1625b I according to \citet{Teachey-etal-2018}, region in cyan is the range proposed by \citet{Martin-etal-2019} and region in green is the range of mass from $0.75$ to $1.25$ $M_{Nep}$.

For all cases we can see that there is a high concentration of smaller satellites inside $10$ $R_p$, close to our inner boundary at $5$~$R_p$, usually these bodies have a more massive companion spread outside. Also, independent of the initial size of the disc, in all cases we found satellites forming within the regions proposed by \citet{Teachey-etal-2018} and by \citet{Martin-etal-2019} for Kepler 1625b I, including bodies with masses in the range of mass expected for Kepler 1625b I. Curiously, for the setup \texttt{kepler-087} the stability limit is inside the region proposed in \citet{Martin-etal-2019}, since satellites formed outside the stability region are expected to not survive long time due to the presence of the star.

In our stability study, we found the same length for the outer stability region in star-planet separation of $1.5$~au and $2$~au, however analysing panels \texttt{kepler-15} and \texttt{kepler-2} of Figs. \ref{fig:kepler-a-m} we can see a clear difference in the satellite population after $100$~$R_p$. While for setup \texttt{kepler-15} there is drastic decrease in the number of surviving satellites in this region, the same is not true for setup \texttt{kepler-2}, where a significant population of relative smaller mass satellite is observed, including one satellite outside the stability limit of $200$~$R_p$. We can see from the lower panels of Fig. \ref{fig:stability} that bodies near the stability limit could have high eccentric orbits, with the case of $a_p = 2$~au even more eccentric than in the case of $a_p = 1.5$~au.

\subsubsection{Eccentricity of the Satellites}

\begin{figure*}
  \begin{center}
  \includegraphics[scale=0.67]{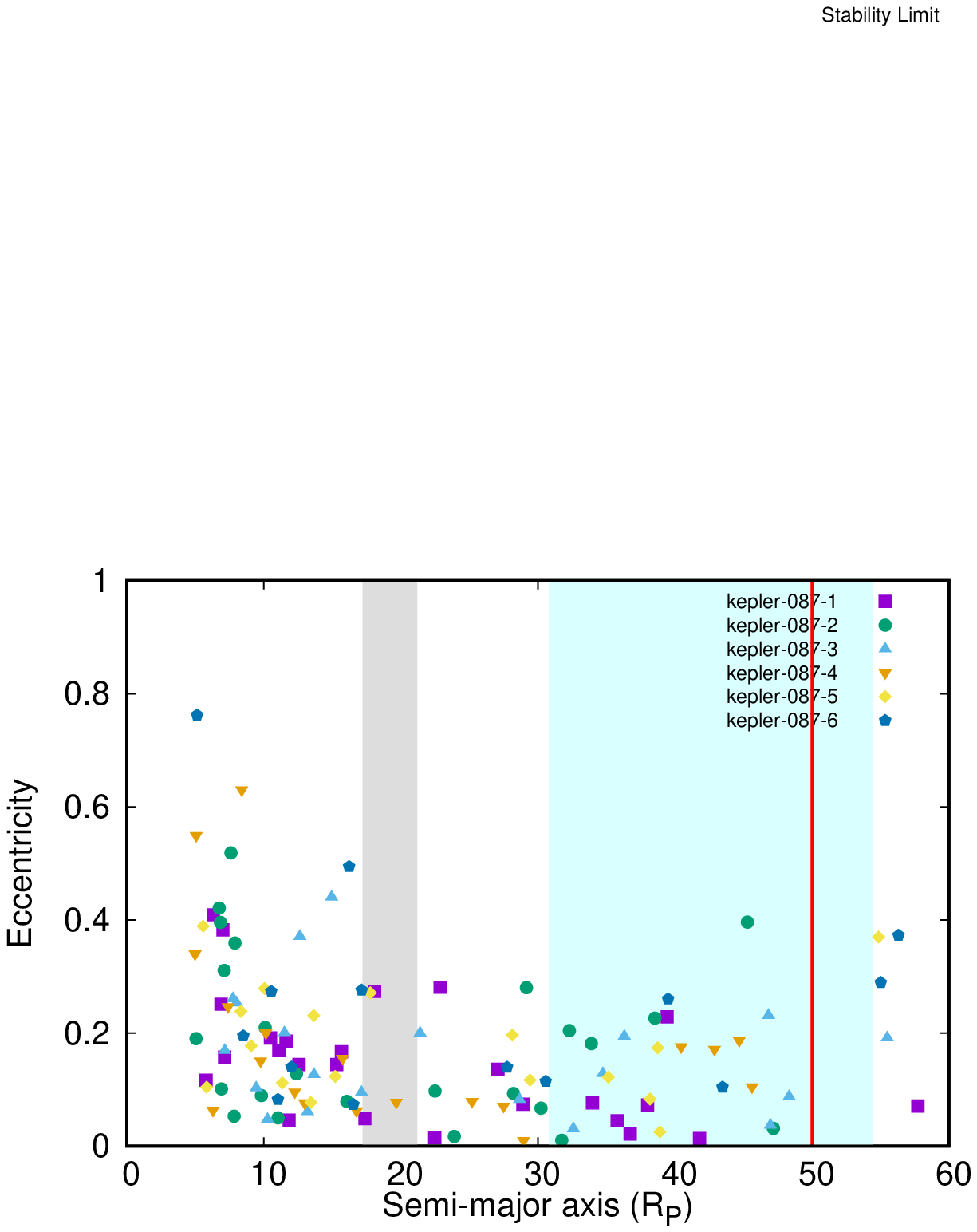}
  \includegraphics[scale=0.67]{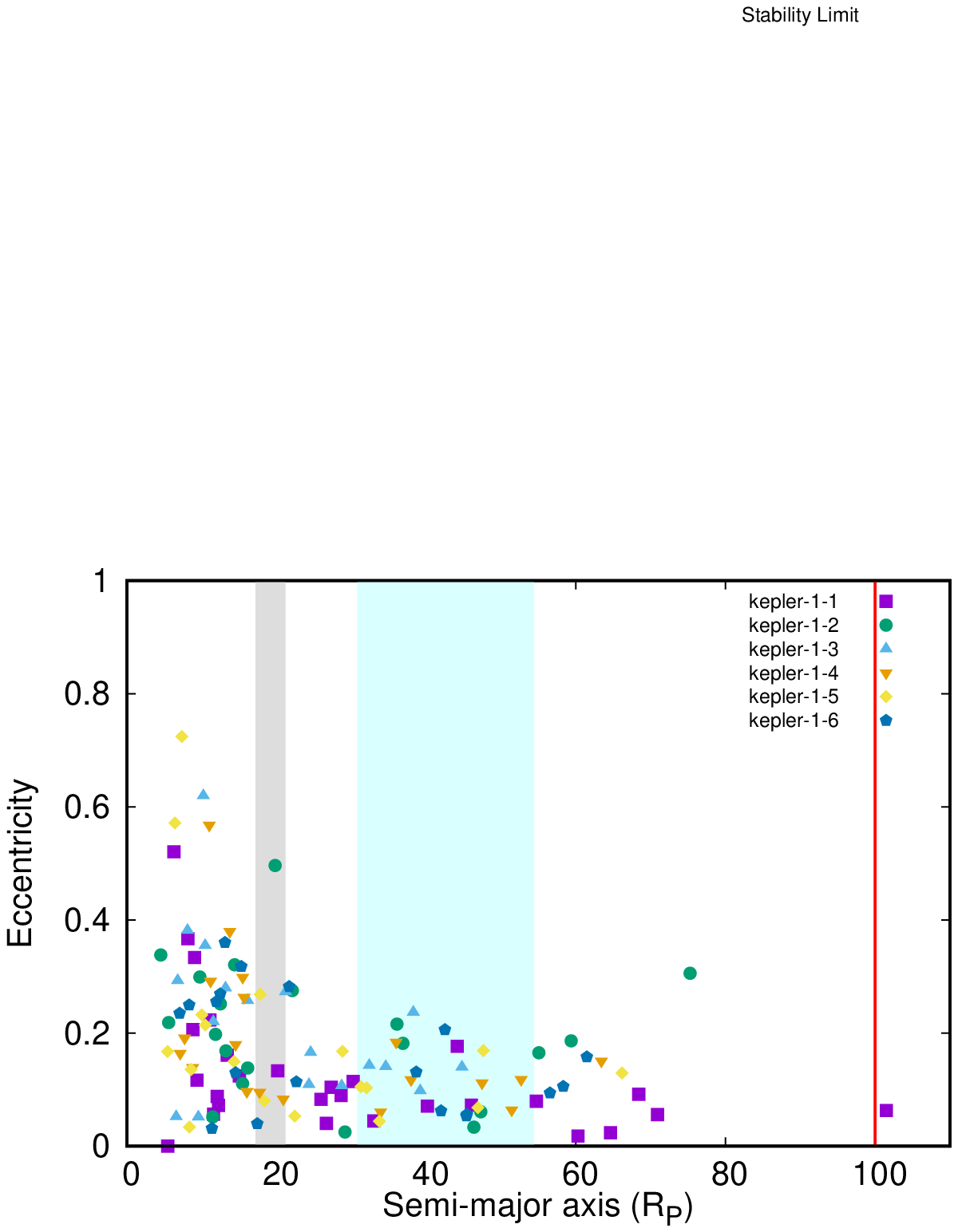}
  
  \includegraphics[scale=0.67]{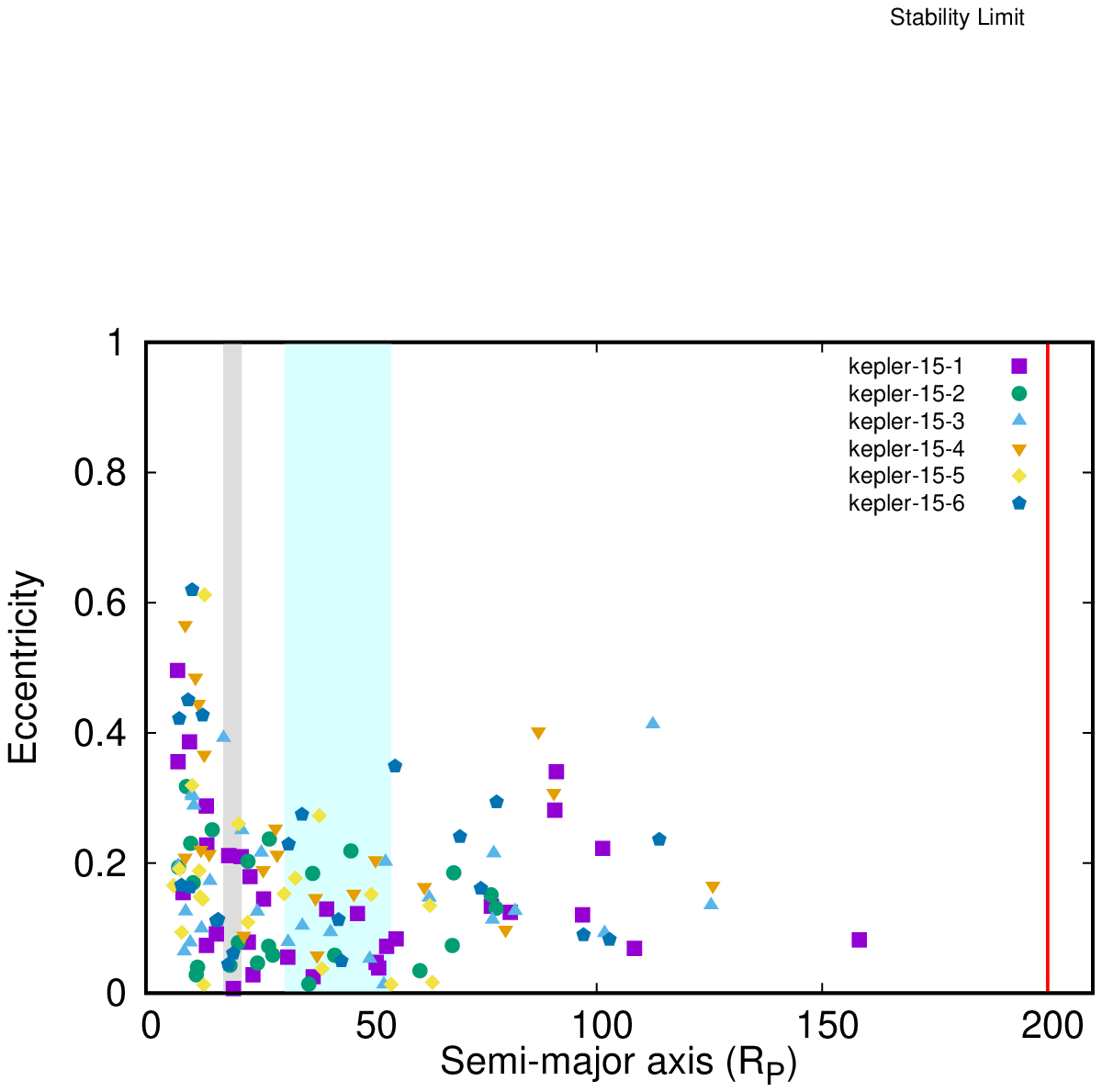}
  \includegraphics[scale=0.67]{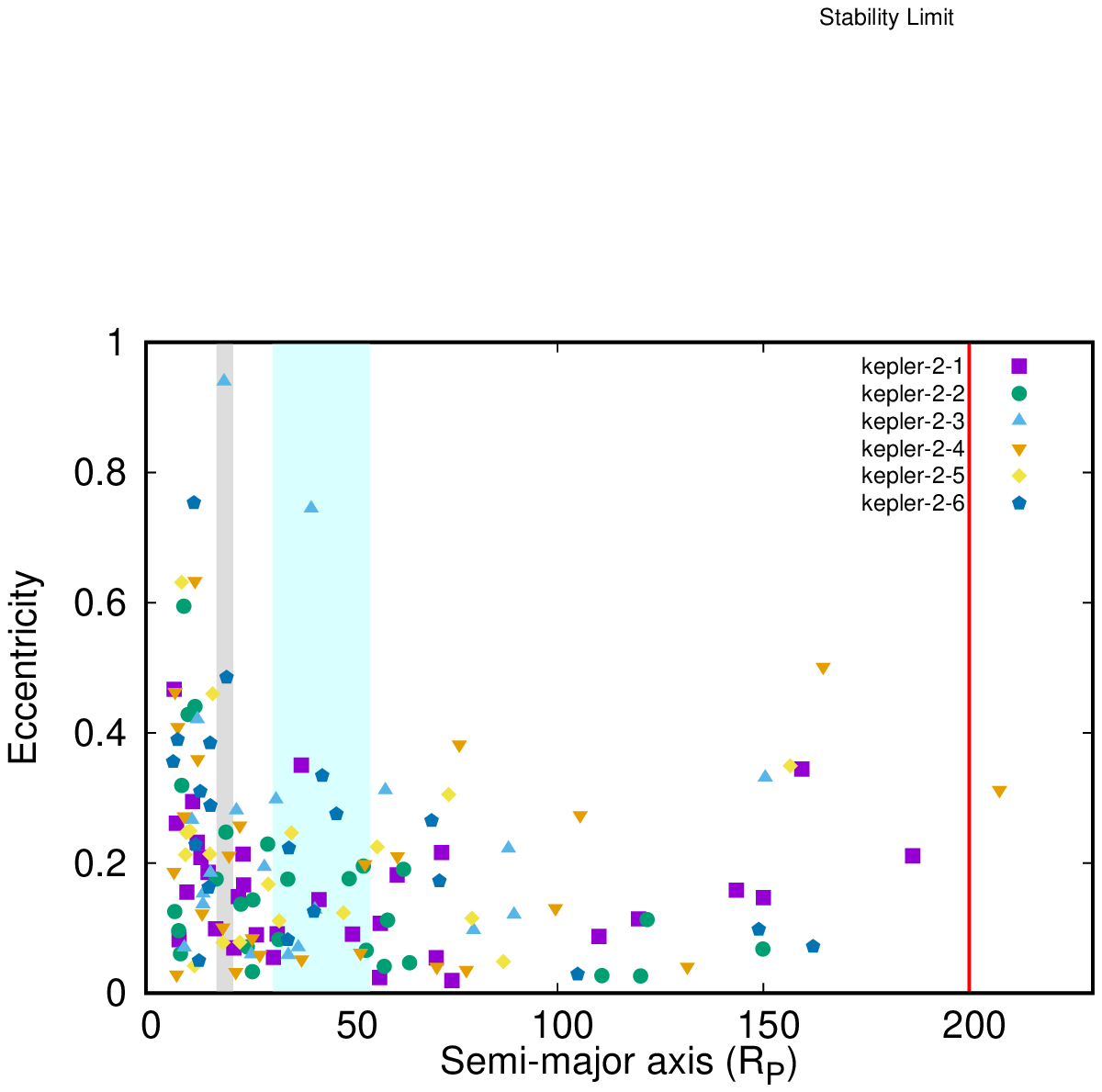}
  \caption[Kepler-ecc1]{Panels showing eccentricity vs semi-major axis for each setup. The region in grey is the location of the exomoon according to \citet{Teachey-etal-2018} while the region in cyan is the range of semi-major axes proposed by \citet{Martin-etal-2019}. The region in green is the range of mass from $0.75$ to $1.25$~$M_{Nep}$. The red line marks the limit of stability obtained in section \ref{stwo}. The formed satellites are separated by the initial mass of the disc: 1~$M_{Nep}$ in purple squares; 2~$M_{Nep}$ in green circles; 3~$M_{Nep}$ in light-blue upward triangles; 4~$M_{Nep}$ in orange downward triangles; 5~$M_{Nep}$ in yellow diamonds; 6~$M_{Nep}$ in dark-blue pentagons.}
  \label{fig:kepler-ecc} 
  \end{center} 
\end{figure*}

From Table \ref{tab:results-formation} one can see that the average eccentricities of the formed satellites are higher than that we find in our Solar System for regular satellites. This outcome is due to the nature of satellites formed by massive bodies confined in tight disc. The satellite embryos underwent a history of energetic collisions and close encounters, most of the bodies were ejected and only a few satellites survived this hostile environment. From the aforementioned table solely, we cannot draw many conclusions or find patterns about the eccentricity of the satellites, thus in Fig. \ref{fig:kepler-ecc} we show the eccentricities of each surviving satellite as a function of its semi-major axis.

The first pattern we observe is that the most eccentric satellites are located near the inner boundary of the system. As the distance from the planet increases the satellite eccentricities decrease, maintaining an average eccentricities around $0.2$, consistent with Table \ref{tab:results-formation}.

Comparting Fig. \ref{fig:stability} with Fig. \ref{fig:kepler-ecc} we see quite an opposite effect. In the stability study, eccentricities were higher close to the stability boundary, while here the eccentricities are higher close to the planet. This result shows that the formation process overcomes the evection effects during the collisional events.

Most of the satellites inside the region proposed by \citet{Martin-etal-2019} for Kepler 1625b I have eccentricities lesser than $0.2$ and could benefit from post-formation damping effects, such as tides, to decrease their eccentricities.

Comparing our satellite systems to multi-planet systems with an external perturber, it is expected that the satellite-satellite interactions and the perturbation coming from the star cause excitation on the eccentricity and inclination of the bodies, specially if no damping effects are considered \citep{Sotiriadis-etal-2017, Pu-Lai-2018}. This would explain the configurations of the surviving satellites in our setups.

\subsubsection{Mass of the Satellites}

The main goal of this work is to set the minimum mass in solids needed in the circum-planetary disc for a $1$ Neptune mass satellite could consistently be formed. We found that the mass of the satellites depends not only on the amount of material in the disc, but also on the size of the disc. As larger the disc initially was, more mass in the disc was necessary to form massive satellites.

\begin{figure}
  \begin{center}
  \includegraphics[scale=0.67]{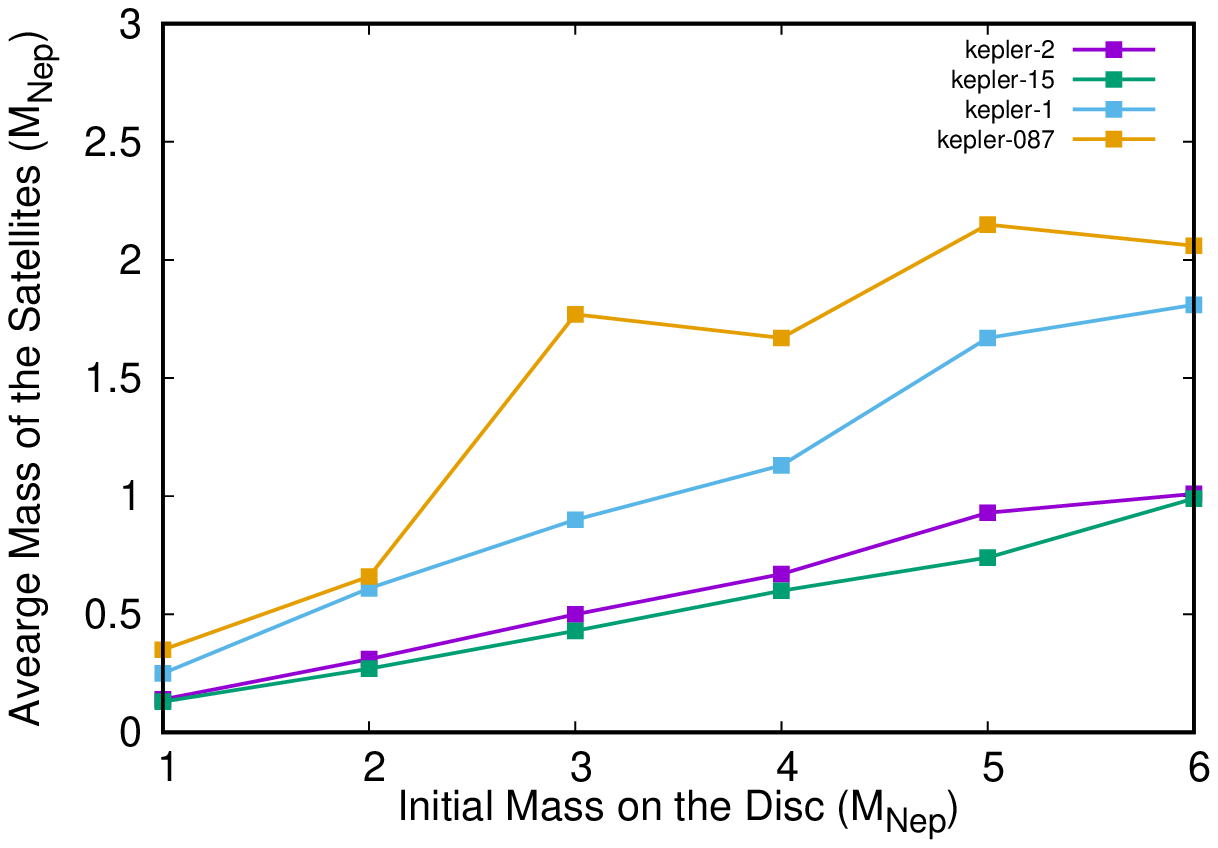}
  \caption[Mass-size]{Correlation between the initial mass in the dis and the average mass of the satellites for models \texttt{kepler-087} (orange line), \texttt{kepler-1} (blue line), \texttt{kepler-15} (green line) and \texttt{kepler-2} (purple line).}
  \label{fig:mass-size} 
  \end{center} 
\end{figure}

The correlation between the average mass of the satellites with the size of the circum-planetary disc and its initial mass is shown in Fig. \ref{fig:mass-size}. The size of the disc is given from the separation setup. For longer discs the distances between embryos and satellitesimals are greater, which leads to less accretive collisions and consequently less massive satellites. We recall that for setup \texttt{kepler-15} and \texttt{kepler-2} the discs have the same initial length, given the similarities shown on the average mass of the satellites.

As expected, in all cases the average mass of the satellites increased as the disc (embryos and satellitesimals) became more massive. Except for setup \texttt{kepler-087} the correspondence between satellite masses and initial mass on the disc was almost linear. The disperse behaviour presented in setup \texttt{kepler-087} is due to the compactness of the formed systems and the presence or absence of small satellites in the inner portion of the disc, which affected the average results.

In Table \ref{tab:results-mass} we summarize our results for the minimum amount of solids necessary for a satellite with around one Neptune mass to form consistently for every different star-planet separation. The results presented are based on Fig. \ref{fig:kepler-a-m}. 

\begin{table*}
\caption[Results Mass]{Results of the simulations regarding the minimum mass to form a satellite with mass comparable to Neptune. Here, ${a}_{p}$, is star-planet separation and $M_{solids}$ is the range of minimum mass composed by solids initially in the disc necessary to form a Neptune-like satellite.}
\begin{tabular}{lcc} \hline
 Model	                & ${a}_{p}$            & $M_{solids}$ \\
                        & au                  & $M_{Nep}$   \\  \hline
\texttt{kepler-087}   & $0.87$               & $2.0 - 3.0$   \\ 
\texttt{kepler-1}     & $1.0$                & $2.0 - 3.0$   \\ 
\texttt{kepler-15}    & $1.5$                & $4.0 - 5.0$\\ 
\texttt{kepler-2}     & $2.0$                & $4.0 - 5.0$\\ \hline

\end{tabular}
\label{tab:results-mass}
\end{table*}

\subsection{Tidal Evolution}\label{subsec:tides}

For the tidal evolution, we selected only systems with at least one satellite with mass between $0.75$ and $1.25$ Neptune mass. Many of them have more than one satellite in the system, and a few of these other satellites have mass greater than one Neptune mass. We separate the results according to the star-planet separation.

The majority of the systems have at least two bodies with the smaller satellite inside $20$ $R_p$. We expect the satellites inside $20$ $R_p$ to be the most affected by tides.

The migration direction due to tidal interaction depends on the corotation radius of the planet. If the satellite orbits within this distance the migration is inward, while when the satellite orbits outside this radius the migration is outward. In this way, we expect the smaller satellites orbiting near the planet to migrate inwardly through tidal interaction with the planet, and eventually collide with the planet. The possibility of tidal disruption of the satellites and the formation of rings \citep{Leinhardt-etal-2012} around Kepler 1625b is beyond the scope of this work.

As the inner satellites migrate inward, their eccentricity and inclinations are also expected to rapidly decrease. For satellites migrating outward, the damping on eccentricity and inclination should be smoother.

\subsubsection{Tidal Evolution: Results}

\begin{figure*}
  \begin{center}
  \includegraphics[width=\textwidth, height=9.5cm]{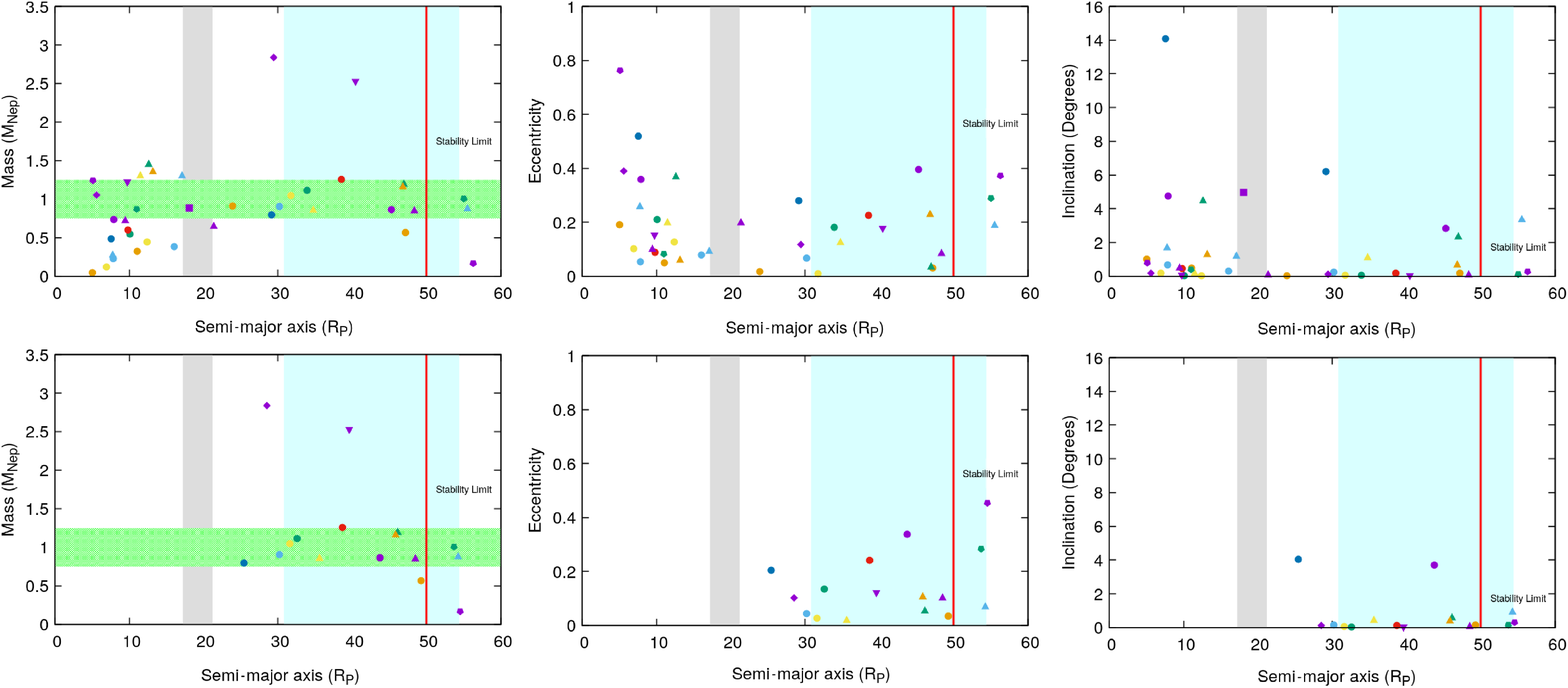}
  \caption[finaltides1]{First row from left to right: Initial radial distribution of mass, eccentricity and inclination of the selected satellites for $a_p = 0.87$~au. Second row from left to right: Radial distribution of mass, eccentricity and inclination of the selected satellites for $a_p$ after 100\,000 years of tidal evolution. Satellites from the same system are indicated by figures with the same shape and colour. The shapes also indicate the type of disc in which the satellite was formed: $1$ $M_{Nep}$: squares; $2$ $M_{Nep}$: circles; $3$ $M_{Nep}$: upward triangles; $4$ $M_{Nep}$: downward triangles; $5$ $M_{Nep}$: diamonds; $6$ $M_{Nep}$: pentagons.}
  \label{fig:tides-out-087} 
  \end{center} 
\end{figure*}

\begin{figure*}
  \begin{center}
  \includegraphics[width=\textwidth, height=9.5cm]{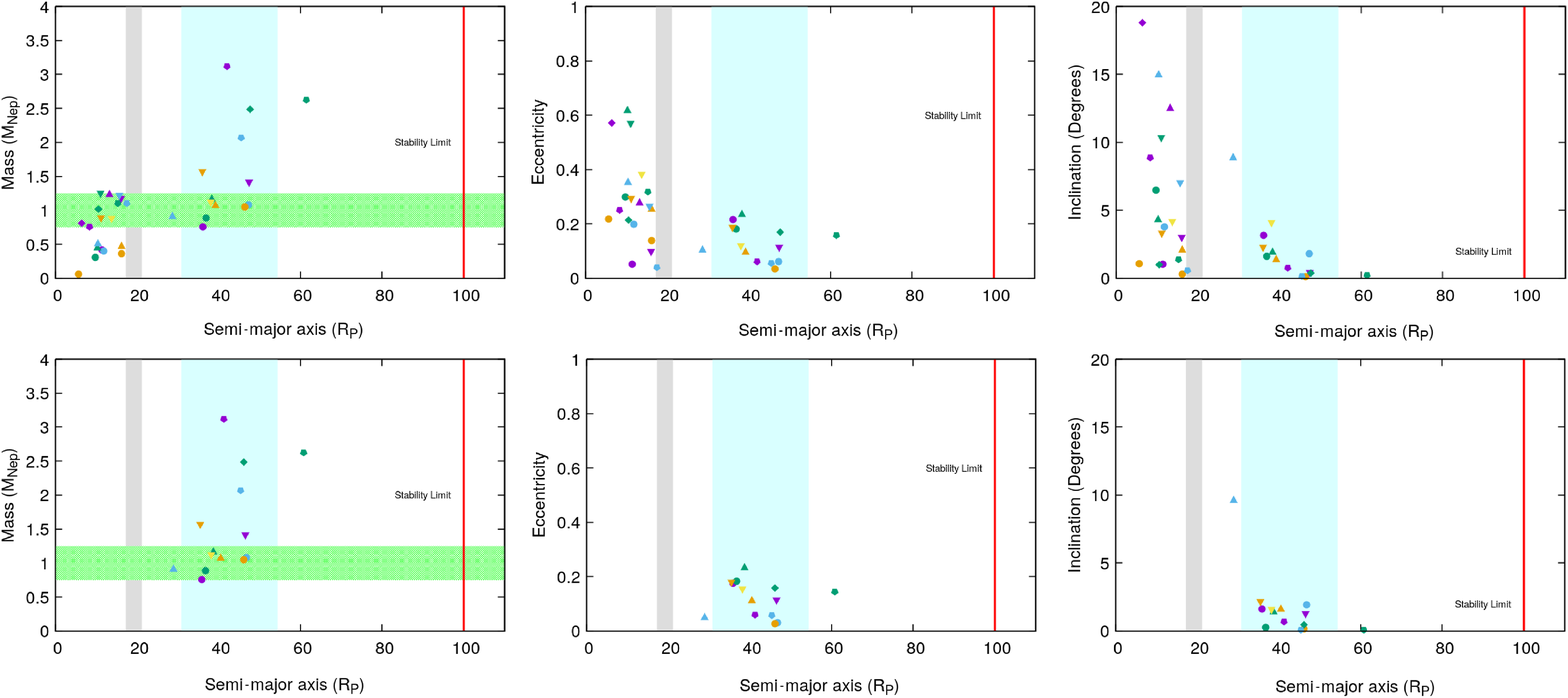}
  \caption[finaltides2]{First row from left to right: Initial radial distribution of mass, eccentricity and inclination of the selected satellites for $a_p = 1.0$~au. Second row from left to right: Radial distribution of mass, eccentricity and inclination of the selected satellites for $a_p$ after 100\,000 years of tidal evolution. Satellites from the same system are indicated by figures with the same shape and colour. The shapes also indicate the type of disc in which the satellite was formed: $1$ $M_{Nep}$: squares; $2$ $M_{Nep}$: circles; $3$ $M_{Nep}$: upward triangles; $4$ $M_{Nep}$: downward triangles; $5$ $M_{Nep}$: diamonds; $6$ $M_{Nep}$: pentagons.}
  \label{fig:tides-out-1}
  \end{center} 
\end{figure*}

\begin{figure*}
  \begin{center}
  \includegraphics[width=\textwidth, height=9.5cm]{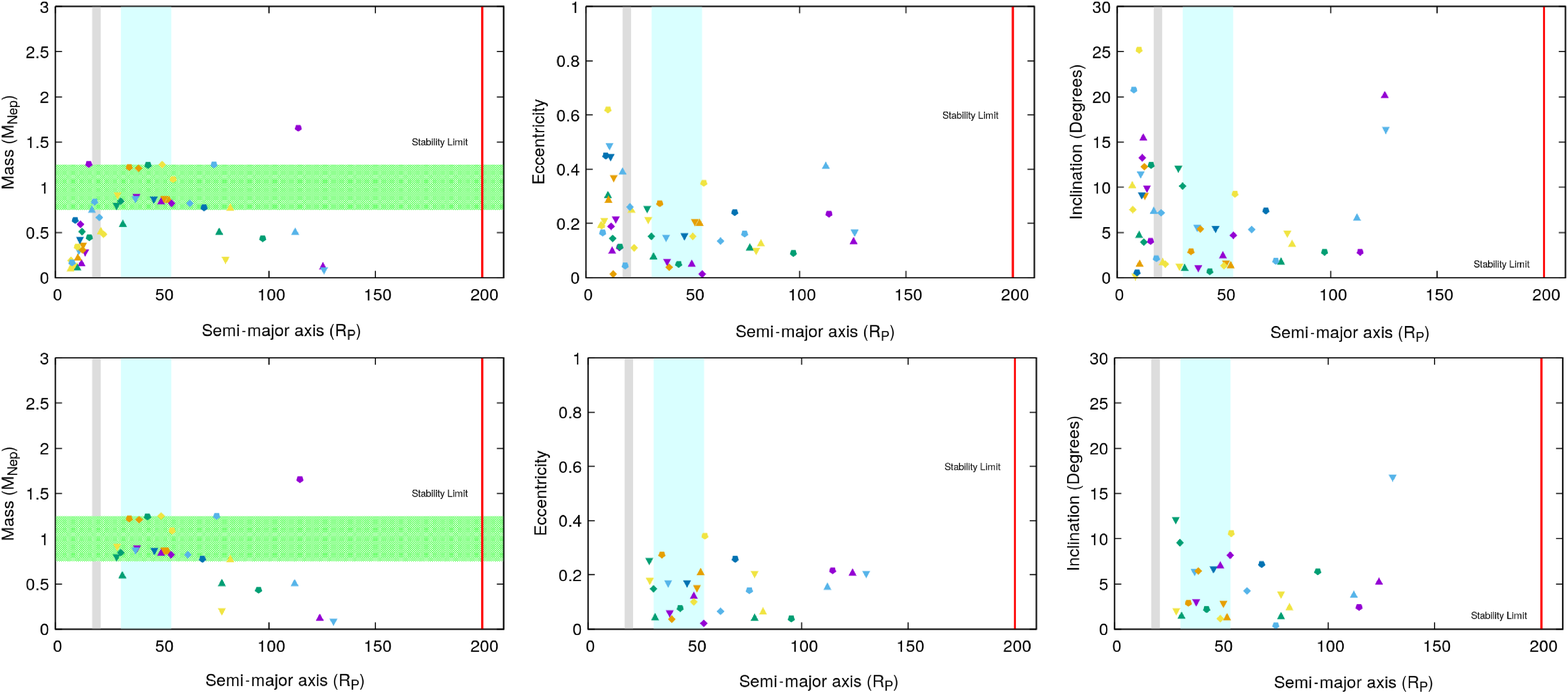}
   \caption[finaltides3]{First row from left to right: Initial radial distribution of mass, eccentricity and inclination of the selected satellites for $a_p = 1.5$~au. Second row from left to right: Radial distribution of mass, eccentricity and inclination of the selected satellites for $a_p$ after 100\,000 years of tidal evolution. Satellites from the same system are indicated by figures with the same shape and colour. The shapes also indicate the type of disc in which the satellite was formed: $1$ $M_{Nep}$: squares; $2$ $M_{Nep}$: circles; $3$ $M_{Nep}$: upward triangles; $4$ $M_{Nep}$: downward triangles; $5$ $M_{Nep}$: diamonds; $6$ $M_{Nep}$: pentagons.}
  \label{fig:tides-out-15} 
  \end{center} 
\end{figure*}

\begin{figure*}
  \begin{center}
  \includegraphics[width=\textwidth, height=9.5cm]{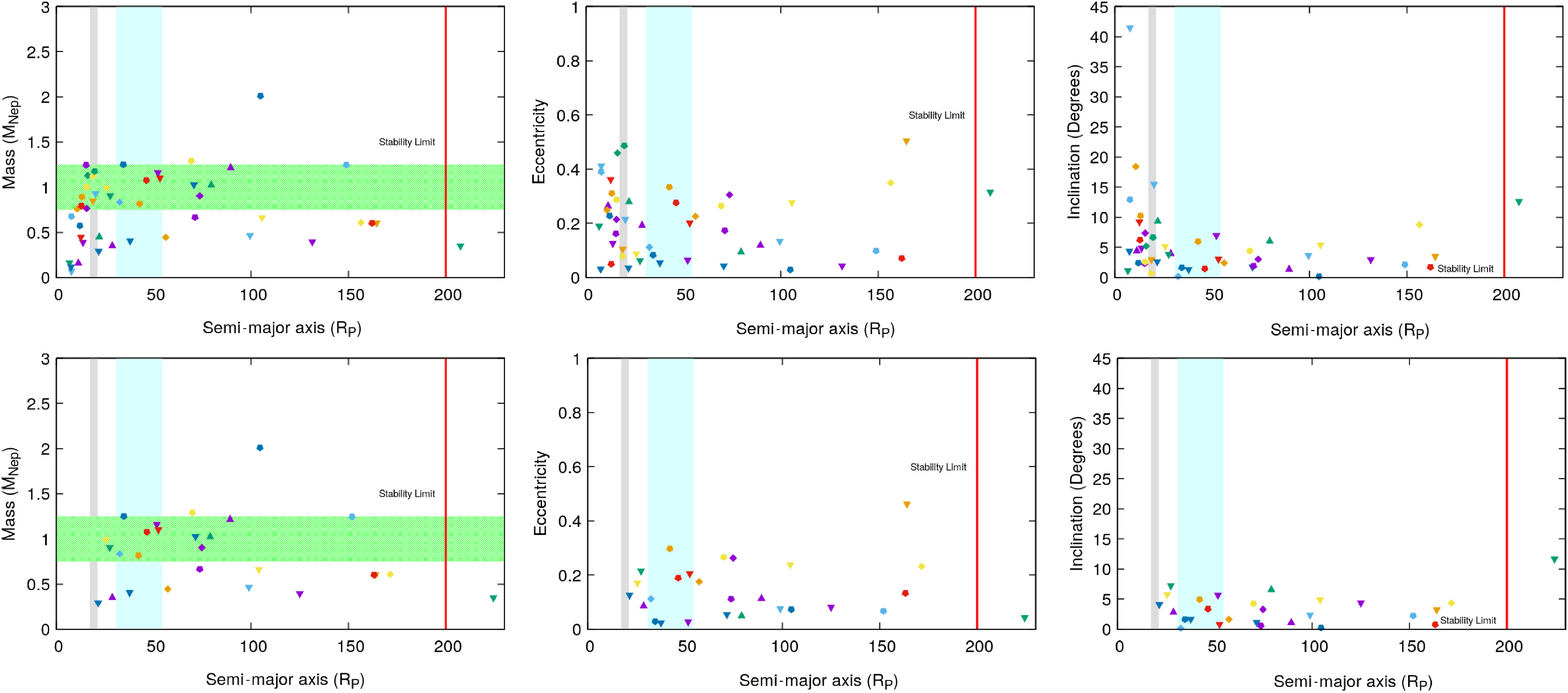}
  \caption[finaltides4]{First row from left to right: Initial radial distribution of mass, eccentricity and inclination of the selected satellites for $a_p=2.0$~au.  Second row from left to right: Radial distribution of mass, eccentricity and inclination of the selected satellites for $a_p$ after 100\,000 years of tidal evolution. Satellites from the same system are indicated by figures with the same shape and colour. The shapes also indicate the type of disc in which the satellite was formed: $1$ $M_{Nep}$: squares; $2$ $M_{Nep}$: circles; $3$ $M_{Nep}$: upward triangles; $4$ $M_{Nep}$: downward triangles; $5$ $M_{Nep}$: diamonds; $6$ $M_{Nep}$: pentagons.}
  \label{fig:tides-out-2}
  \end{center} 
\end{figure*}

The simulations including tides were carried for 100\,000 years. We found that after this time, the effects of tidal effects were not relevant to the orbital evolution of the surviving satellites. In the first row of Figs. \ref{fig:tides-out-087} - \ref{fig:tides-out-2} we show the initial configuration of the selected systems, the second row is the configuration after the simulation. Satellites indicated by figures with the same shape and colour are part of the same system.

As expected, in all scenarios the satellites inside $20$ $R_p$ rapidly migrated inward towards the planet and collided with the central body within 1\,000 years, during this phase they experienced a strong process of circularization in their orbits. The inclinations of these bodies were also affected, such that coplanar configuration was  quickly achieved in the inner disc, before the loss of these satellites.

Beyond $20$ $R_p$ the effects of tidal interaction were more modest. The satellites' migration rate outside this distance is irrelevant compared with the migration of the inner satellites, indicating that they are stable for a long period of time. The tidal effects on the eccentricity and inclination were also weaker, however the average eccentricity and inclination of the satellites substantially decreased, since the most extreme satellites located near the planet were lost during the evolution.

Comparing the predictions of \citet{Teachey-etal-2018} (grey region) and \citet{Martin-etal-2019} (cyan region) for a possible location of Kepler 1625b I, our simulations indicate the former as preferable, since, in all cases, the satellites initially inside the region proposed by \citet{Teachey-etal-2018} were lost. Also, most of the surviving satellites with masses comparable to the mass of Neptune were located inside the region proposed in \citet{Martin-etal-2019}, with the best result for 1~au separation.

\section{Conclusions}\label{conc}
\label{conclusion} 

In this work we have investigated the necessary conditions to form a Neptune-like satellite in system Kepler 1625. This system is composed by a star, a giant planet and a satellite candidate around the planet. Due to many uncertainties regarding the physical and orbital characteristics of the system, we opted to study satellite formation using four different star-planet separations, $0.87$~au, $1$~au, $1.5$~au and $2$~au, in this way we covered scenarios of satellite formation in different stages of the planet's evolution. We also considered the two possible locations for the exomoon candidate, the one proposed by \citet{Teachey-etal-2018} and the one proposed by \citet{Martin-etal-2019}. For tidal analysis we verify a broad range of masses, $1.0{\pm0.25}$ $M_{Nep}$. The expected characteristics of the Kepler 1625 system are shown in table \ref{tab:properties}.

For our purposes we used dedicated N-body simulations using the numerical package MERCURY divided in three phases: a) stability studies of particles around Kepler 1625b; b) satellite formation; c) post-formation tidal interactions between satellites and planet. As it was shown by \citet{Domingos-etal-2006}, a ten Jupiter mass planet locate at 0.87 au from the star, such as Kepler 1625b, could harbour a Neptune-mass satellite in prograde motion inside 0.4985 Hill radius of the planet. Also, this upper limit coincides with the lower boundary of the evection resonance. In our study, we distributed 10\,000 massless particles inside of one Hill radius of the planet considering different semi-major axes for the planet. As expected, the stability region is highly dependent on the star-planet separation, with the star playing a decisive role over the final stability limit, the closer is the star, the smaller is the region of stability. Our results regarding stability are shown in Fig. \ref{fig:stability}, and the values for the outer stability border were used in table \ref{tab:formation} where they were rounded up in order to cover more extreme cases. From Fig. \ref{fig:stability} we can see that for $a_p = 0.87$~au, particles can be stable only within $0.22$ $R_H$, well inside the distance where evection resonance is predicted to be dominant. In fact, except for the case with $a_p = 1.5$~au (where particles were stable up to $0.51278$ $R_H$ during the 10\,000 years integration), we see that particles orbiting beyond $0.5$ $R_H$ are scattered from the systems, due to evection resonances. Also, at the outer border of stability we have regions where high eccentric particles can survive, probables reminiscent of the scattered particles.

After finding the stability regions for the systems, we simulated discs with satellite embryos and satellitesimals and analyse the growth and orbital evolution of the satellite candidates. The sizes of the discs were based on our results regarding stability. In this way, we have four discs with different sizes, according to the star-planet separation. For each different disc we simulated six different setups with 1 to 6 Neptune mass in total. The number of embryos and satellitesimals are the same in all cases, however their masses are scaled according to the model (Table \ref{tab:formation}). To minimize the effects of the initial distribution, for each amount of mass ten different initial distributions were applied to the satellite embryos, all setups used the same distribution scale law and different initial masses \citep{Kokubo-Ida-2002, Raymond-etal-2005}.

In all cases, the majority of the systems formed harbour more than one satellite, sometimes even four. This outcome is due to  many reasons, such as, initial mass of the bodies, separation between them and disc size. We draw a correlation relating the initial mass of the embryos and the number of formed satellites, we found that the number of surviving bodies tends to decrease in discs with initially high massive embryos (see column two of table \ref{tab:results-formation}). In these cases, the close encounters between two bodies are more energetic and often ends with one of the embryos being ejected from the system. Also, we found that, in average, systems with two satellites are preferable, with one body inside $20$ $R_p$ and the other in a radially wider orbit. 

\citet{Teachey-etal-2018} and \citet{Martin-etal-2019} predicted two different location for the exomoon candidate Kepler 1625b I. Because of the discrepancy in these two predictions we compared our results with both. In all cases, there is a high concentration of all sorts of satellites in the inner disc, inside $20$ $R_p$, however, these satellites are not expected to survive a post-formation evolution due to tidal interactions since these effects tend to be stronger for closer satellites to the planet. In this case, we can say that the prediction made by \citet{Martin-etal-2019} seems to be more likely. Moreover, we found the concentration of surviving bodies to decrease near the stability limit, this is due to the star presence. A few satellites managed to survive outside the stability limit of their respective system, but we consider these cases to be exceptions.

Satellites formed in-situ are thought to have almost circular and coplanar orbits with respect to the planet due to their interactions with the circum-planetary disc. However, from our results (Table \ref{tab:results-formation} and Fig. \ref{fig:kepler-ecc}), we found this might not be the case for Kepler 1625b I. In average, the satellites formed with eccentricities bigger than $0.15$ and average inclinations varying from $\sim1^{\circ}$ to $\sim9^{\circ}$, with the most inclined satellite having almost $25^{\circ}$. The majority of these high eccentric high inclined bodies are orbiting near the planet, in this region the gravitational interactions between the satellites and the planet are strong and the perturbation on the orbit of the satellite is expected. Also, our formation scenario is placed after the dissipation of the gas disc around Kepler 1625b, when the satellite embryos are already formed, in this way no damping effects are considered at this point. Recalling that tidal damping was applied after this initial phase in only a handful of selected systems. Near the stability limit of each system, bodies with eccentric and inclined orbit were also expected, yet the sample of surviving satellites located at this specific region is so small that no conclusions can be drawn.

Our main goal was to find a minimum amount of mass in solids for the disc, for a satellite massive as the planet Neptune could form consistently, in this was we tested discs with six different total masses from $1$ to $6$ $M_{Nep}$. From Fig. \ref{fig:kepler-a-m} one can have the full picture of our simulations regarding mass and position of the formed satellites. Analysing these figures we have that the green region on the graphs (range of mass $1.0{\pm0.25}$ $M_{Nep}$) is well populated by satellites formed in almost all discs, meaning that formation of stable massive moons is possible in the Kepler 1625 system. As expected, when the initial mass of the embryos and satellitesimals increases the average mass of the formed satellites also increases. From Fig. \ref{fig:mass-size} we can see that the correlation between initial mass in the disc and average mass of satellites is almost linear. Also, we found that for setups \texttt{kepler-087} and \texttt{kepler-1} discs initially with $2.0 - 3.0$ $M_{Nep}$ in solids are capable to produce satellites with mass comparable to Neptune consistently. The same type of satellites was formed in models \texttt{kepler-15} and \texttt{kepler-2}, however the initial amount of solids necessary had to be increased to $4.0 - 5.0$ $M_{Nep}$, as summarized in Table \ref{tab:results-mass}. 

We used the mass of the formed satellites as the criteria to select the systems to underwent tidal evolution. In order to try to reproduce the satellite system proposed for the Kepler 1625 system, we select all systems with at least one satellite with mass $1.0{\pm0.25}$ $M_{Nep}$, ignoring its location.

We found that the tidal evolution plays a key role in shaping the final architecture of the satellite systems. In our first analysis we found several satellites surviving close to the planet. However, once these bodies were inside the corotation radius of the planet they migrated inward and rapidly collided with the central body. From all our selected systems, the tidal interactions with the planet were responsible for emptying the region inside $20$ $R_p$, consequently almost all the systems ended only with one satellite agreeing with the perspectives for the Kepler 1625 system. Also, after tidal evolution, most of the satellites within the range of expected mass for Kepler 1625b I were orbiting inside the region proposed \citet{Martin-etal-2019}, which seems to be the more likely location of the exomoon candidate.

The tidal evolution was not an effective mechanism for satellites beyond $20$ $R_p$, for these cases  we found that satellites slightly migrated (inward or outward). Also, the damping on eccentricity and inclination were not enough to lead bodies to circular and/or coplanar configurations.

The exomoon candidate Kepler 1625b I does not fit the characteristics of an in-situ formed satellite when compared to satellites of our Solar System, mainly because of its mass. Two constraints against this hypothesis are: the satellite-planet mass ratio law observed in our Solar System, which lies in the order $10^{-4}$ between regular satellites and its host planet; and the amount of mass on the circum-planetary disc, specially solid material. Here we extrapolated the idea of an enhanced-massive circum-planetary disc to find a range of mass in the disc in which a Neptune-like satellite will be formed consistently. According to our models, we found that the amount of solids necessary to form such a massive satellite depends on the semi-major axis of the planet and the size of the region of stability. Also, the hypothesis of a system where more than one satellite can not be neglected, our results showed that the formation of a multi-satellite system around Kepler 1625b is possible. In addition, we demonstrated the importance of taking into account a post-formation tidal evolution of the satellite systems. With simulations including tides we were capable to reproduce the system of satellites expected for Kepler 1625b, in mass and semi-major axis, from different initial conditions. However, the eccentricity and inclination of the formed satellites was still not compatible with what we expect for in-situ formed bodies for our Solar System.

In addition to the studies performed by \citet{Heller-2018} and \citet{Hamers-Zwart-2018} we expected to give other possible explanation for the origin of the exomoon candidate Kepler 1625b I. We argue that in-situ formation cannot be neglected only based on the patterns observed in our Solar System. We have shown that, given an initial mass in the disc, Neptune-like satellites could form around Kepler 1625b. More information about the exomoon candidate such as the orbital direction of motion will provide more pieces for this puzzle and accurately point towards a scenario of origin for Kepler 1625b I.

\section*{Acknowledgements}
We thank the anonymous referee for the valuable comments and suggestions and Muller Lopes and Barbara Camargo for the computational help. RAM and EVN thanks financial support from FAPESP (Grant: 2011/08171-3).





\bibliographystyle{mnras}
\bibliography{ref}

\appendix
\bsp	
\label{lastpage}
\end{document}